\documentclass{jfm}
\usepackage{lineno,hyperref}
\usepackage{graphicx}
\usepackage{floatrow}
\usepackage{epstopdf, epsfig}
\usepackage{upgreek}
\usepackage{subfigure}
\usepackage{amsmath}
\usepackage{amssymb}
\usepackage{mathrsfs}  
\usepackage{nomencl}
\usepackage{setspace}
\usepackage[round]{natbib}
\usepackage[usenames,dvipsnames]{xcolor}
\usepackage{tikz}
\usetikzlibrary{shapes}
\usepackage{MnSymbol}
\usepackage[normalem]{ulem}
\usepackage[ruled,vlined]{algorithm2e}
\usepackage{wrapfig}
\usepackage{colortbl}
\usepackage{mathtools}
\usepackage{xfrac}

\shorttitle{Synchronization of turbulence in channel flow}
\shortauthor{M. Wang and T. A. Zaki}

\def\m1line{\vrule width3pt height2.5pt depth -2pt}

\def\bdot{\raise.2em\hbox to .15em{.}}

\usepackage{todonotes}
\title{Synchronization of turbulence \\ in channel flow}

\author{
Mengze Wang
\and  Tamer A.~Zaki
\corresp{\email{t.zaki@jhu.edu}}
}

\affiliation{Department of Mechanical Engineering, 
Johns Hopkins University,
Baltimore, MD 21218, USA}

\begin{document}

\maketitle

\begin{abstract}

Synchronization of turbulence in channel flow is investigated using continuous data assimilation.  
The flow is unknown within a region of the channel. 
Beyond this region the velocity field is provided, and is directly prescribed in the simulation, while the pressure is unknown throughout the entire domain.  
Synchronization takes place when the simulation recovers the full true state of the flow, or in other words when the missing region is accurately re-established, spontaneously.  
Successful synchronization depends on the orientation, location and size of the missing layer.  
For friction Reynolds numbers up to one thousand, wall-attached horizontal layers can synchronize as long as their thickness is less than approximately thirty wall units. 
When the horizontal layer is detached from the wall, the critical thickness increases with height and is proportional to the local wall-normal Taylor microscale.  
A flow-parallel, vertical layer that spans the height of the channel synchronizes when its spanwise width is on the order of the near-wall Taylor microscale, while the criterion for a crossflow vertical layer is set by the advection distance within a Lyapunov timescale.   
Finally, we demonstrate that synchronization is possible when only planar velocity data are available, rather than the full outer state, as long as the unknown region satisfies the condition for synchronization in one direction.  These numerical experiments demonstrate the capacity of accurately reconstructing, or synchronizing, the missing scales of turbulence from observations, using continuous data assimilation. 

\end{abstract}

%

\section{Introduction}
Accurate predictions of turbulence must contend with its chaotic and multiscale nature: A slight deviation in the initial or boundary conditions exponentially amplifies and leads to an inaccurate prediction of all scales, and hence signifcant deviations of the trajectories in state space \citep{Deissler1986chaotic,Nikitin2018}.
In addition, it is difficult to precisely measure all the scales of turbulence, especially at high Reynolds numbers due to the larger separation of scales. 
Data assimilation methods aim to reconstruct the full state of a dynamical system from limited observations \citep{EnKFbook,DAbook}.  These methods provide estimated trajectories that shadow the true state within an observation time horizon and that forecast the flow evolution beyond the observations, although, naturally, with progressively decreasing accuracy due to chaos.
In homogeneous isotropic turbulence, recent studies have demonstrated that observing all the scales that are larger than about twenty Kolmogorov lengthscales is sufficient to reconstruct the missing smaller eddies, accurately\textemdash a phenomenon termed synchronization of chaos \citep{Yoshida2005,Eyink2013}. In this work, we investigate synchronization in turbulent channel flow, in physical rather than spectral space.  Specifically, we report on the maximum size of an cloaked region that can be synchronized to the true flow trajectory using outer observations, and the required observations.

In order to incorporate observations into nonlinear dynamical systems, three types of data assimilation approaches have been developed: variational methods \citep{Dimet1986_4dvar,Li2020,Wang_hasegawa_zaki_2019}, ensemble methods \citep{EnKFbook,Mons2016}, and continuous data assimilation techniques \citep{Charney1969ds,Yoshida2005}.
Variational methods seek the optimal initial condition that minimizes a cost function, defined in terms of the distance between the estimated trajectory and available data.
The minimization procedure requires the gradient of the cost function, which is efficiently computed using the adjoint equations.  
Ensemble methods do not involve an adjoint model. They consist of a prediction step where an ensemble of states are advanced using the forward equations, and an analysis step which assimilates observations to update the estimated flow state. 
The two classes can be combined, where an ensemble of forward simulations is adopted to construct a quadratic approximation of the cost function and to evaluate the gradient, which is generally referred as the ensemble variational method \citep{Liu2008envar,Mons2019}. 
All of these techniques are widely adopted by the numerical weather prediction community and have been applied to estimating turbulent systems.
However, due to the high-dimensional nature of turbulence, these methods are computationally expensive, require numerous, costly numerical simulations to obtain converged results \citep{Wang2019,Jahanbakhshi_zaki_2019,Buchta2021envar}, which prohibits their application for synchronization of the estimated state with the true trajectory, especially at high Reynolds numbers. 


Continuous data assimilation, which includes nudging techniques \citep{Hoke1976nudging,Lakshmivarahan2013nudging}, augments the governing equations with a forcing term that drives the estimation towards available observations.
Therefore, only one forward numerical simulation is required, which is the lowest cost compared with other approaches.  Accurate predictions depend on the forcing scheme and, most importantly, the available observations. 
Construction of the forcing term is referred to as the ``observer problem'' in control theory:
Given a dynamical system, the ``observer'' refers to a forced system that assimilates limited data with the objective of synchronizing to the original system \citep{Huijberts2001observer}. 
Although different forcing strategies have been developed \citep{Nijmeijer1998observer,Mohan2017observer}, they have a relatively minor influence on the success of synchronization in comparison to the amount of available observations of the true state. 
Therefore, we adopt the most straightforward continuous data assimilation approach: direct substitution, where we replace part of the state vector by the corresponding observation data.
This approach does not depend on any artificial parameters, and thus we can focus on the effect of observation on synchronization. 

Recent efforts in homogeneous isotropic turbulence \citep{Yoshida2005,Eyink2013,Martin_2021} demonstrated that observations of the velocity field at all scales $k \eta < 0.2$,  where $k$ is the wavenumber and $\eta$ is the Kolmogorov lengthscale, can guarantee synchronization of the smaller scales.  
No previous study has investigated synchronization of chaos in wall-bounded turbulence, where the criteria for synchronization should be inhomogeneous due to the no-slip boundary condition and mean shear.  We will therefore perform continuous data assimilation using direct substitution in physical space, and report on the maximum volume of turbulence that can be cloaked and yet synchronizes to the true trajectory by aid of observations of the rest of the system.

It is helpful to contrast synchronization in physical space to the previous studies that were performed in spectral space.  There, the turbulence was prescribed at all wavenumbers above a cutoff value, and hence observations were available at any spatio-temporal location for all the energy-containing and inertial scales.  Synchronization  of a cloaked sub-volume in physical space is markedly different because within the cloaked region, and for all times, none of the flow scales are known. Therefore, it is more difficult to anticipate whether the turbulence can be accurately generated from observations of the true state outside the sub-volume.  
Consider, for example, removing a wall-attached horizontal layer in turbulent channel flow.  This region includes the vorticity flux at the wall, which generates all the interior vorticity \citep{Lighthill,Wu2007book,Eyink2020_theory}, and therefore synchronization using only outer observations may be difficult to achieve.  If the layer includes the region of peak turbulence production, again whether it is possible to synchronize this ``engine" of turbulence is uncertain. 
Non-local interactions, for example of outer large-scale motions with near-wall structures \citep{Marusic2009LSM, Bernardini2011, hwang_2016}, may promote synchronization. Previous efforts have remarked on the role of these interactions in state estimate using linear \citep{Baars2016,Illingworth2018} or nonlinear approaches \citep{Sasaki_2019,Wang2021}, but none have examined synchronization. 
When the unknown layer is vertical and normal to the mean-flow direciton, streamwise advection of upstream observations can aid synchronization.  If the unknown vertical layer is, however, parallel to the flow, synchronization over the entire channel height is anticipated to be relatively more challenging.  In addition, it is difficult to anticipate the amount of external observations that are required for synchronization to take place.  
All these considerations will herein be examined in the context of turbulent channel flow, for the first time.

\begin{figure}
	\centering
	\includegraphics[width=\textwidth]{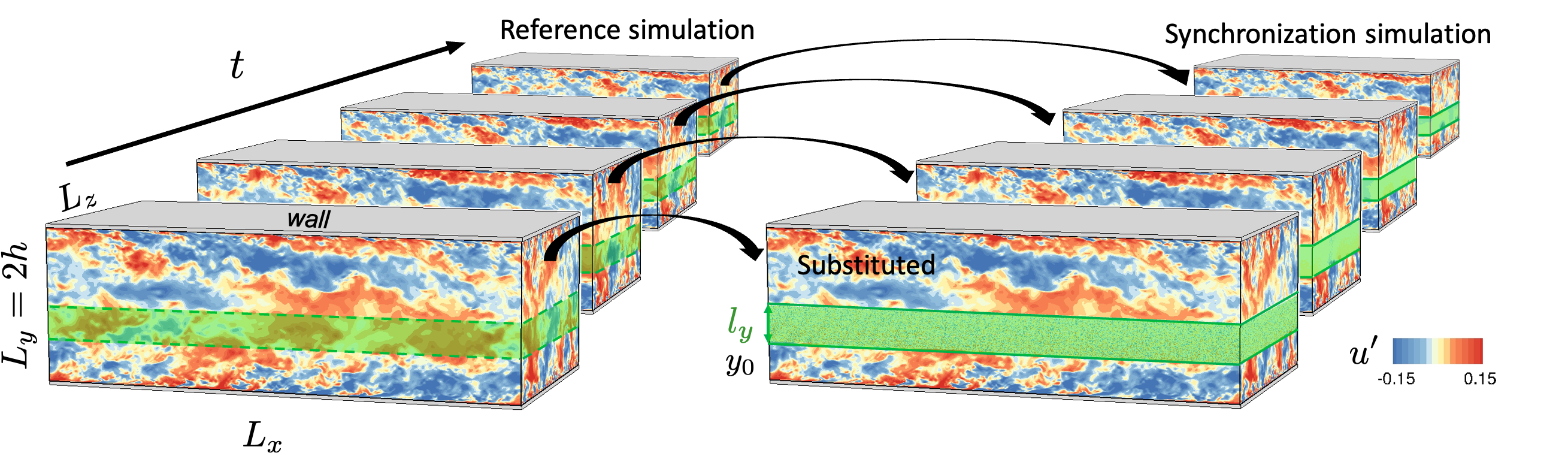}
	\caption{Schematic of the reference and synchronization simulations of turbulent channel flow. 
			A sample cloaked, or unobserved, horizontal region in the synchronization simulation is marked in green.}
	\label{fig:setup}
\end{figure}

In \S\ref{sec:methods}, we provide details of the computational setup, and introduce the direct substitution algorithm.
We apply the method in \S\ref{sec:wall_layer} to synchronize a horizontal wall-attached layer by observing the fully-resolved outer flow, and we report the maximum layer thickness for successful synchronization and the influence on Reynolds number.
Synchronization of wall-detached and of multiple layers are examined in \S\ref{sec:log_layer} and \S\ref{sec:multiple}, respectively, and the equation for the evolution of synchronization error is derive in appendix \ref{sec:error_eqn}.
In \S\ref{sec:subdomain}, we assess synchronization of a flow region using surface observations only, and discuss potential ways to improve the estimation accuracy of this approach.  Additional tests for sub-domain synchronization in Kolmogorov flow are presented in appendix \ref{sec:Kflow}.
Concluding remarks are provided in \S\ref{sec:conclusion}.

\section{Computational approach}
\label{sec:methods}

The reference flow configuration is a rectangular channel (see figure \ref{fig:setup}), which is periodic in the streamwise ($x$) and spanwise ($z$) directions, and bounded by two parallel no-slip surfaces in the vertical direction ($y$). The channel half height $h^*$ and bulk flow velocity $U_b^*$ are adopted as the reference length and velocity scales, where $^*$ denotes dimensional quantities.  When quantities are scaled by inner variables, specifically the friction velocity $u_{\tau}^*$ and kinematic viscosity $\nu^*$, they are marked by superscript $^+$.
The bulk and friction Reynolds numbers are $Re \equiv U_b^* h^*/\nu^*$ and $Re_{\tau} \equiv u_{\tau}^*h^*/\nu^*$, respectively.

Our objective is to perfectly reconstruct the unknown turbulent field within a region $\Omega_s$ of the channel, where we do not have any data, from observations of the outer flow in a region $\Omega_f$.  For example, in figure \ref{fig:setup} the horizontal green layer $y \in [y_0,y_0+l_y]$ is the cloaked region $\Omega_s$ and observations are available in the entire outer domain $\Omega_f$.  The full data assimilation domain is $\Omega = \Omega_s \cup \Omega_f$.  Perfect reconstruction of $\Omega_s$ here means synchronization to the true flow that generated the outer observations, to within machine precision.  
We start by introducing the numerical schemes, computational setup, and synchronization algorithm.


The velocity field $\boldsymbol u$ and pressure $p$ satisfy the incompressible Navier-Stokes (N-S) equations, 
\begin{eqnarray}
\label{eq:cont_div}
\boldsymbol{\nabla} \cdot \boldsymbol{u} &=& 0  \\
\label{eq:cont_mom}
\frac{\partial \boldsymbol{u} }{\partial t} +  \boldsymbol{\nabla} \cdot (\boldsymbol{uu}) &=& - \boldsymbol{\nabla} p + \frac{1}{Re} \nabla^2 \boldsymbol{u} 
\end{eqnarray}
where $t$ represents time. 
These equations are spatially discretized on a staggered grid with a local volume-flux formulation \citep{Rosenfeld1991}, and advanced using a second-order fractional-step method \citep{Moin_fractional,dpform} that adopts Adams-Bashforth scheme for the advection terms and Crank-Nicolson for diffusion.
Due to periodicity in $x$ and $z$, the pressure Poisson equation is solved using Fourier transform in these directions and tri-diagonal inversion in $y$ direction. 
The algorithm has been validated and applied in a number of direct numerical simulations (DNS) of transitional and turbulent flows 
\citep{Zaki2013,Marxen_2019}.

The true, or reference, system $\boldsymbol u_r(t)$ is an equilibrium turbulent flow that is sustained by a constant pressure gradient in the streamwise direction.
The domain sizes and grid resolutions for all the considered Reynolds numbers are summarized in table \ref{table:Re}.  The two highlighted configurations, at $Re_{\tau}=\{590, 1000\}$, will be the focus of the majority of the discussion.  The influence of domain size is examined at $Re_{\tau}=590$, where a larger domain $(L_x,L_z) = (8\pi, 3\pi)$ is also considered in order to accommodate very-large-scale structures \citep{delAlamo2004,Choi2004}. 
The time step size $\Delta t$ in each case is chosen to guarantee that the Courant–Friedrichs–Lewy (CFL) number is less than one half.


\begin{table}
    \centering
    \begin{tabular}{c c c c c c c c c c c}
        \hline
	    \multicolumn{2}{c}{Parameters} & \multicolumn{2}{c}{Domain size}  & \multicolumn{3}{c}{Grid points} & \multicolumn{4}{c}{Grid resolution} \\
    	$Re_{\tau}$ & $Re$ & $L_x/h$ & $L_z/h$ & $N_x$ & $N_y$ & $N_z$ & $\Delta x^+$  & $\Delta y^+_{min}$ & $\Delta y^+_{max}$ & $\Delta z^+$ \\
    	\hline
    	180 &  2,800 & $4\pi$ & $2\pi$ &  ~~384 &  256 &  ~~320 &  5.9 &  0.20 & 3.0 & 3.5 \\
    	392 & 6,875 & $2\pi$ & $\pi$ & ~~256 & 320 & ~~192 & 9.6 & 0.34 & 5.1 & 6.4 \\
    	\rowcolor{blue!10}
    	590 & 10,935 & $2\pi$ & $\pi$ & ~~384 & 384 & ~~384 & 9.7 & 0.44 & 6.5 & 4.8 \\
       	590 & 10,935 & $3\pi$ & $\pi$ & ~~576 & 384 & ~~384 & 9.7 & 0.44 & 6.5 & 4.8 \\
       	590 & 10,935 & $4\pi$ & $\pi$ & ~~768 & 384 & ~~384 & 9.7 & 0.44 & 6.5 & 4.8 \\
    	590 & 10,935 & $8\pi$ & $3\pi$ & 1,536 & 384 & 1,152 & 9.7 & 0.44 & 6.5 & 4.8 \\
    	\rowcolor{blue!10}
    	1,000 & 20,000 & $2\pi$ & $\pi$ & ~~768 & 768 & ~~768 & 8.2 & 0.29 & 5.8 & 4.1 \\
    	\hline
    \end{tabular}
    \caption{Computational domains and grid sizes for simulations at different Reynolds numbers. 
    }
    \label{table:Re}
\end{table}

The observer system, or synchronization simulation, is denoted $\boldsymbol u_s(t)$ and is
performed in the domain $\Omega = \Omega_s \cup \Omega_f$ which, unless otherwise state, is the same as the reference simulation.
Given the fully-resolved true velocity field $\boldsymbol u_r(t)$ in sub-volume $\Omega_f$, these data are directly enforced onto the synchronization simulation, $\boldsymbol{u}_s=\boldsymbol{u}_r~\forall \boldsymbol{x}\in\Omega_f$ and $t$. 
The reference pressure field $p_r(t)$ is not observed, and hence the pressure in the synchronization simulation $p_s(t)$ differs from the true state throughout the entire domain $\Omega$.
Infusing the velocity observations into the synchronization simulation is performed using the direct substitution algorithm \ref{alg:ds}, which is also illustrated in figure \ref{fig:setup}.  Due to the non-local nature of the Navier-Stokes solution, errors within $\Omega_s$ instantaneously contaminate the solution within $\Omega_f$, and therefore the direct substitution procedure is enforced every time step.  
Once applied, $\boldsymbol u_s$ differs from $\boldsymbol u_r$ only within the unobserved region $\Omega_s$, where the synchronization error is defined,
\begin{equation}
    \label{eq:err_V}
    \mathcal E_{s} \equiv \langle \|\boldsymbol u_s - \boldsymbol u_r \|^2 \rangle_{\Omega_s}^{1/2},
\end{equation}
where $\| \boldsymbol a\|$ is the 2-norm of vector $\boldsymbol a$, and $\langle \bullet \rangle_{\Omega_s}$ denotes averaging over $\Omega_s$.
Theoretically, synchronization refers to  
\begin{equation}
    \label{eq:sync}
    \lim_{t \to \infty} \langle \|\boldsymbol u_s - \boldsymbol u_r \|^2 \rangle_{\Omega_s}^{1/2} = 0,
\end{equation}
for any arbitrary choice of initial condition $\boldsymbol u_s(0)$ within $\Omega_s$.
The condition (\ref{eq:sync}) also guarantees that the pressure field $p_s$, which is not observed, converges to the reference state $p_r$ throughout the entire domain.
Due to finite numerical precision and integration time, the above condition will be approximated by 
\begin{equation}
    \mathcal E_s(t) < \beta \quad \text{when} \ t < T,
\end{equation}
The threshold $\beta$ is set to $10^{-15}$ due to our double-precision arithmetic, and $T=80$ is sufficiently long to achieve a conclusive trend.

The initial condition $\boldsymbol u_s(t=0)$ for $\boldsymbol{x}\in \Omega_s$ is estimated using one of three approaches:
(i) a trivial initial condition $\boldsymbol{u}_s(t=0) = \boldsymbol{0}$;
(ii) the local mean velocity superposed with white noise proportional to the local root-mean-square fluctuations, $\boldsymbol u_s = \langle \boldsymbol{u}_r\rangle + \boldsymbol{\eta}$, where $\eta_i \sim N(0,u_{i,rms})$ is a Gaussian random field;
(iii) a slight perturbation to the true state, $\boldsymbol u_s = \boldsymbol u_r + \epsilon \boldsymbol \eta$, where $\epsilon = 10^{-4}$.
When synchronization occurs, all these initial conditions lead to exponentially decreasing estimation error with the same rate, which is equal to the leading Lyapunov exponent of the sub-dynamics within $\Omega_s$.
If $\Omega_s$ exceeds a certain size or if the available observations within $\Omega_f$ are insufficient, synchronization is not possible. In such cases, even simulations starting from the initial estimate (iii), which is infinitesimally close to the true state, generate exponentially amplifying errors relative to the reference trajectory. 
Simulations of type (iii) are initially infinitesimally close to the true state, and hence are designed to probe the linear exponential stability of the sub-system; these computations will be termed Lyapunov experiments.



\begin{algorithm}[h]
	\SetAlgoLined
	At initial time $t=0$, directly enforce $\boldsymbol{u}_s(0)=\boldsymbol{u}_r(0)$ in the observed region $\Omega_f$\;
	Estimate $\boldsymbol u_s(0)$ within the unknown region $\Omega_s$\;
	\While{$ \mathcal E_s > \beta$ \and \ $t<T$}{
	Advance the synchronization simulation from $\boldsymbol u_s(t)$ to $\boldsymbol u_s(t+\Delta t)$ using Navier-Stokes (\ref{eq:cont_div}-\ref{eq:cont_mom})\;
	Enforce $\boldsymbol u_s(t+\Delta t) = \boldsymbol u_r(t+\Delta t) ~\forall \boldsymbol{x} \in \Omega_f$\;
    Evaluate $\mathcal E_s$ \;
	$t \leftarrow t + \Delta t$
	}
	\caption{Direct substitution algorithm.}
	\label{alg:ds}
\end{algorithm}



\begin{figure}
	\centering
	\includegraphics[width=\textwidth]{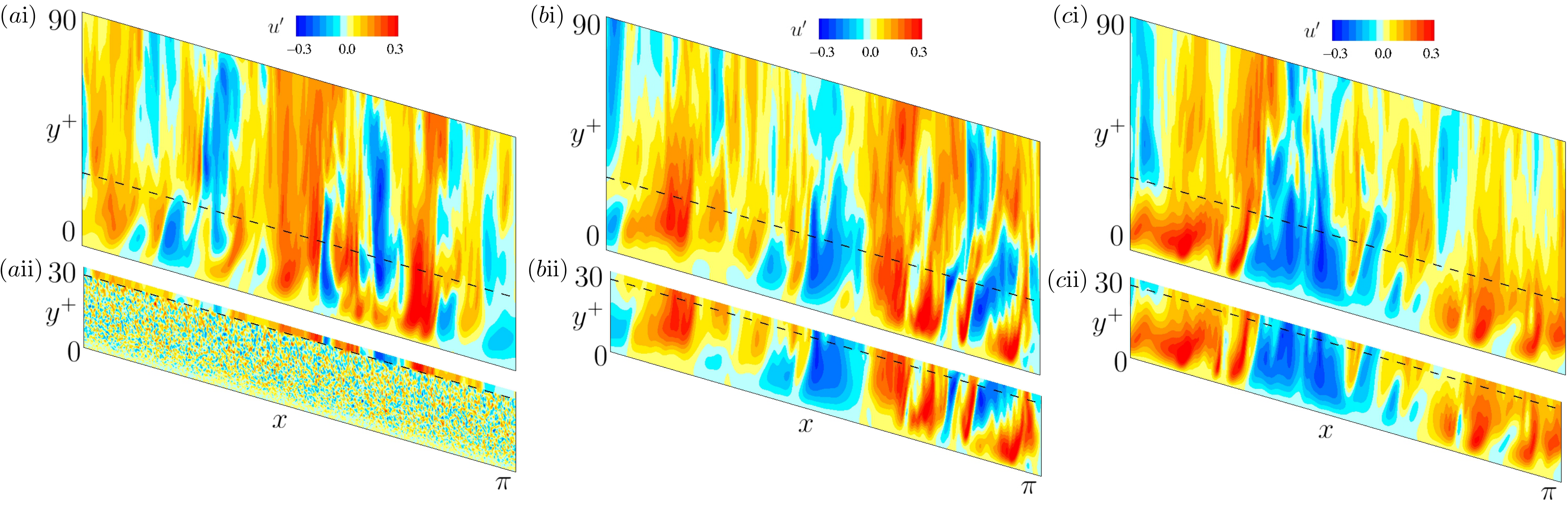}
	\caption{
		Synchronization of a horizontal, wall-attached layer at $Re_{\tau}=1{,}000$. 
		Contours are the instantaneous streamwise velocity fluctuations, calculated by subtracting the true mean, at $z = L_z/2$. 
		($a$-$c$) $t^+=\{ 0, 40, 160\}$; 
		(i) true state; (ii) synchronization simulation. 
		Dashed line: $y^+ = 28$.
	}
	\label{fig:quality}
\end{figure}

\section{Results}
\label{sec:results}
\subsection{Synchronization of a horizontal wall-attached layer}
\label{sec:wall_layer}


We start by attempting to synchronize a wall layer, $\Omega_s = \{ \boldsymbol x \in [0,L_x] \times [0,l_y] \times [0,L_z]\}$, similar to the setup in figure \ref{fig:setup} with the lower boundary on the wall.  
In the limit $l_y\rightarrow 0$, synchronization is essentially trivial.  As $l_y$ increases, successful synchronization becomes less inevitable: The absence of the true vorticity source at the wall from the observer simulation may impede synchronization, while the coupling of the outer observations within $\Omega_f$ to the near-wall region may aid in the convergence of the simulation to the reference trajectory.  Ultimately the balance of restoring and destabilizing effects is anticipated to lead to divergence of the sub-dynamics from the reference simulation, which in the limit of large $l_y$ is again not surprising since the entire streamwise and spanwise dimensions are eliminated, and all scales/wavenumbers in those dimensions at a given height are unknown.  The interesting aspect of this problem is therefore not the limiting behaviours, but whether there exists a specific, or critical, value of $l_y$ across which the behaviour changes and how it depends on Reynolds number.  

\begin{figure}
	\centering
		\includegraphics[width=\textwidth]{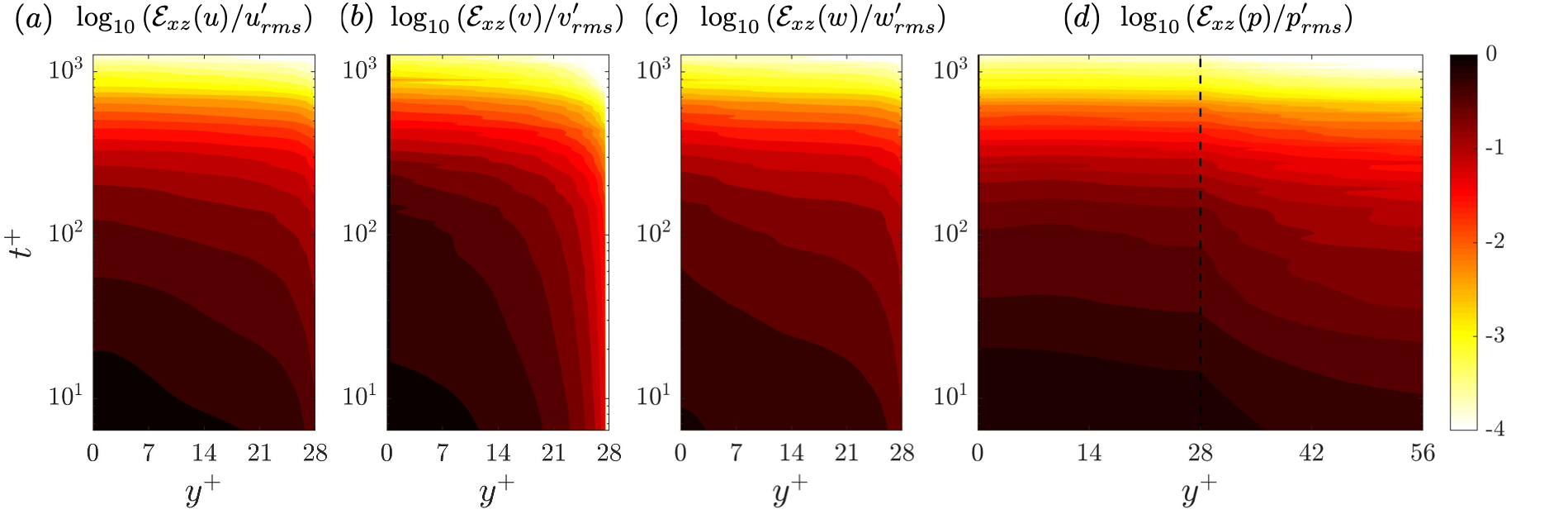}
	\caption{
		Time and wall-normal dependence of the synchronization error.  The error is averaged in the horizontal plane and normalized by the local root-mean-square fluctuation, $\log_{10} \left(\mathcal E_{xz}(q)/q^{\prime}_{rms} \right)$.  ($a$-$d$) $q=\{u, v, w, p\}$.
		}
	\label{fig:err_xz}
\end{figure}

\begin{figure}
	\centering
		\includegraphics[width=0.5\textwidth]{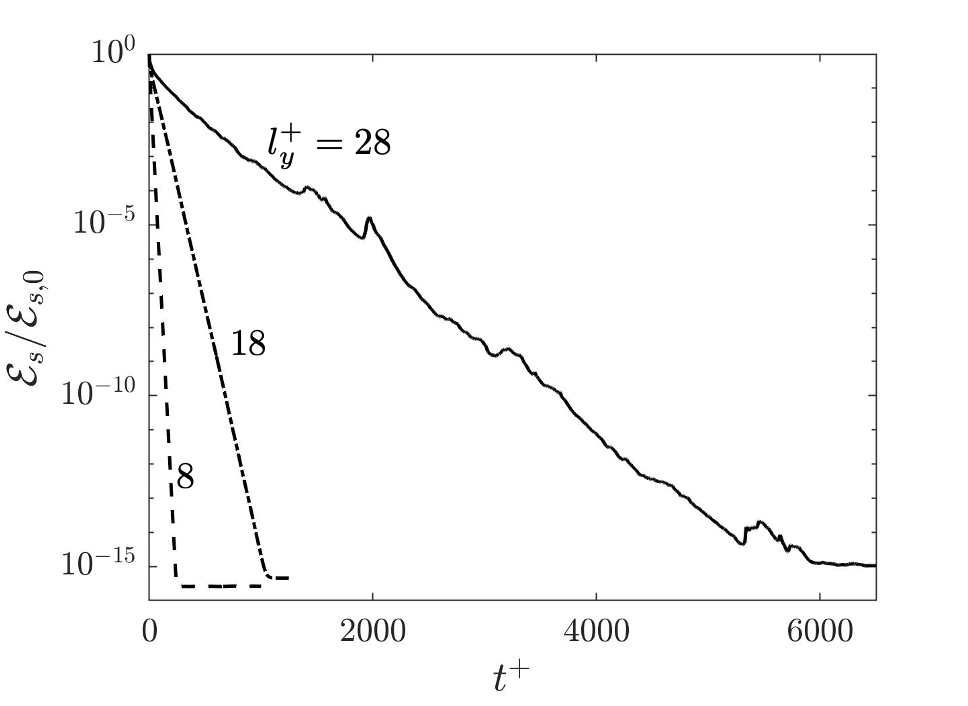}
	\caption{
		Temporal dependence of the volume-averaged synchronization error $\mathcal E_s$ normalized by the value at the initial value $\mathcal E_{s,0}$.  
		Results are for $Re=1{,}000$:
		(dashed line) $l_y^+ = 8$; (dashed dot line) $l_y^+=18$; (solid line) $l_y^+=28$.
	}
	\label{fig:err_V}
\end{figure}

A qualitative account of synchronization at $Re_{\tau}=1{,}000$ is provided in figure \ref{fig:quality}, when $l_y^+ = 28$.  The figure shows a side view of the channel, with the top panels displaying the reference simulation and the bottom panels focused only on the cloaked region $y \in \Omega_s$ of the synchronization simulation. 
Panel (aii) is the initial estimate of $\boldsymbol{u}_s(t=0)$ within $\Omega_s$, which is the mean flow plus white noise.
Synchronization of the wall layer proceeds in two stages:  During an initial transient (panel bii), the white noise decays, and the velocity field in the vicinity of the outer observations becomes qualitatively more accurate.  However, quantitative accuracy is not yet achieved especially in the near-wall region.
In the limit of long time (panel cii), the entire wall layer is accurately reconstructed.
To quantify the accuracy, we introduce the horizontally averaged error,
\begin{equation}
	\label{eq:err_xz}
	\mathcal E_{xz}(q) = \langle \left( q_s - q_{r} \right)^2 \rangle_{xz}^{1/2}, \quad q = u, v, w, p.
\end{equation}
At $t=0$, the error $\mathcal E_{xz}(q)$ is approximately $\sqrt{2}$ of the root-mean-square fluctuations for all three components, because the superposed white noise in $\boldsymbol u_s(0)$ is uncorrelated with the true fluctuations.
The temporal evolution of $\mathcal{E}_{xz}$ is plotted in figure \ref{fig:err_xz}, normalized by the local root-mean-square fluctuations.
Unlike the velocity (panels $a$-$c$) where the errors vanish in $\Omega_f$, the pressure is not observed and hence has finite error for $y^+ > 28$ (panel $d$).
The two stages of error decay are evidenced in figure \ref{fig:err_xz}.
At the initial stage $t^+ \sim O(10)$, the error remains highest near $y^+ = 0$ due to the delayed impact by observations.
Beyond $t^+ = 100$, the synchronization errors within the entire removed layer are diminishing for both velocity and pressure, and uniformly tend to zero.
As such, it is reasonable to focus on the volume-averaged error $\mathcal E_s$ (solid line in figure \ref{fig:err_V}), which decays until it reaches $10^{-15}$ dictated by our double-precision implementation.


\begin{figure}
	\centering
	\includegraphics[width=0.45\textwidth]{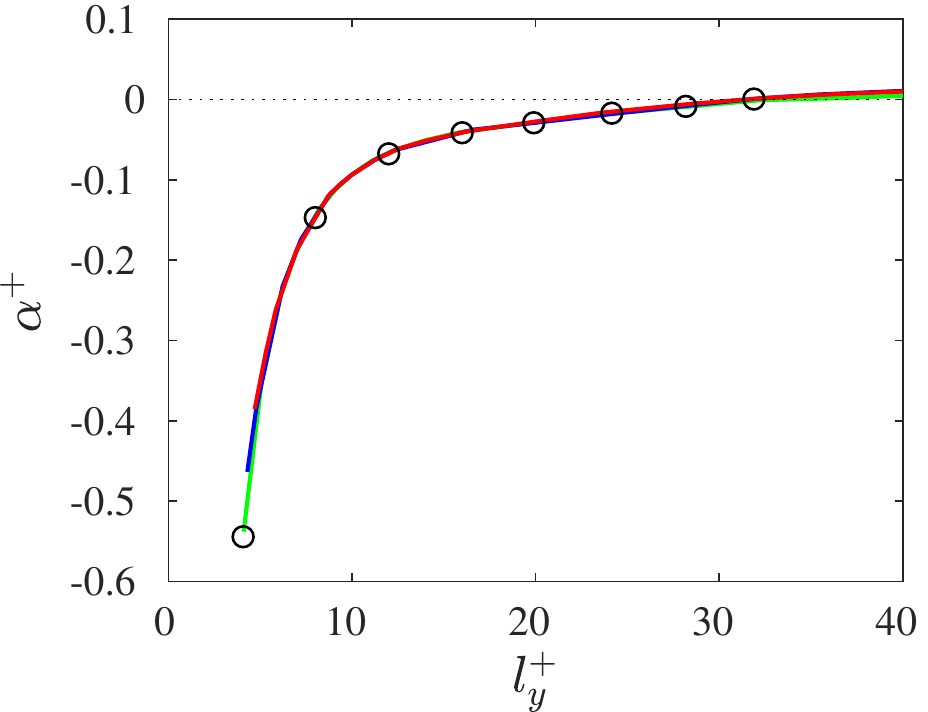}
	\caption{
		Dependence of the synchronization exponents $\alpha^+$ on the thickness $l_y^+$ of the cloaked wall-attached layer, at different Reynolds numbers.
		(Green) $Re_{\tau}=180$; (blue) $Re_{\tau}=392$; (red) $Re_{\tau}=590$;
		(black circles) $Re_{\tau} = 1{,}000$. The dotted line marks $\alpha = 0$.
	}
	\label{fig:slope_Re}
\end{figure}

As the thickness of unknown layer $l_y$ is reduced, synchronization is precipitated at a faster rate, as shown in figure \ref{fig:err_V}.
The rate is evaluated from $\mathcal E_s(t)$ using an exponential regression,
\begin{equation}
    \label{eq:regression}
	\mathcal E_s = A \exp(\alpha t) = A \exp(\alpha^+ t^+),
\end{equation}
where we have introduced the synchronization exponent $\alpha$.  
Only data within the range $10^{-1} \le \mathcal E_{s}/ \mathcal E_{s,0} \le 10^{-6}$ are used for the regression (\ref{eq:regression}) in order to eliminate the transient effects at the beginning of simulations.
The dependence of $\alpha$ on $l_y$ and on Reynolds number is summarized in figure \ref{fig:slope_Re}.
The results demonstrate that indeed as $l_y$ increases, the synchronization rate reduces and ultimately becomes positive beyond a critical value. Note that, for the diverging $\alpha > 0$ cases, the exponent was  determined using the Lyapunov-type experiments where the initial estimate $\boldsymbol{u}_s(t=0)$ was infinitesimally close to the true solution within the cloaked region $\Omega_s$ (case iii, as explained in \S\ref{sec:methods}).

When scaled with viscous units, the profile of $\alpha^+(l_y^+)$ is essentially independent of the Reynolds number up to $Re_{\tau} = 1000$. 
The critical value for $l_y$ that corresponds to zero growth rate, 
\begin{equation}
    \label{eq:yc}
    \alpha(l_{y,c}) = 0,
\end{equation}
is $l_{y,c}^+ \approx 32$, which is the maximum thickness of the wall layer that can be removed without disrupting successful synchronization.

The quantitative relation in figure \ref{fig:slope_Re} is robust against the initial condition of the reference simulation $\boldsymbol u_r(0)$, the initial estimate of $\boldsymbol u_s(0)$ within the cloaked region $\Omega_s$, and the global domain size.
In particular, when the global domain size is increased from $(L_x,L_z)=(2\pi,\pi)$ to $(L_x,L_z)=(8\pi,3\pi)$ at $Re_{\tau}=590$, the value of the synchronization exponent for $l_y^+ = 18$ remains unchanged to within less than $1\%$.  Therefore, synchronization of the wall layer is unaffected by the associated change in the very-large-scale structures in the outer flow \citep{Marusic_perry_1995,Marusic2019review}.

The critical thickness $l_{y,c}^+ \approx 32$ demonstrates that the dynamics within the viscous sublayer and buffer layer are interpretable from outer observations. Even the instantaneous flux of vorticity at the wall is accurately reconstructed by enforcing the fully-resolved outer flow field.  Our results also provide a new perspective on the influence of outer large-scale structures on near-wall turbulence \citep{delAlamo2004,Marusic2009LSM}.
Previous studies established that the near-wall autonomous cycle of streaks and streamwise vortical structures \citep{Jimenez1999,Jimenez2007} is modulated by the coherent structures in the log layer; 
Our synchronization study demonstrates that the instantaneous near-wall cycle can be perfectly regenerated if the complete information about outer turbulence is available. 
Finally, we emphasize that $y^+ \leq 32$ is a diminishing small physical region as $Re$ is increased, which is indicative of an increasing difficulty of synchronization.
The scales of turbulence that are commensurate with the critical thickness are discussed in the next section, where we examine the impact of placing the cloaked horizontal layer at different wall-normal heights.

\subsection{Synchronization of a wall-detached layer}
\label{sec:log_layer}

When the cloaked layer $\Omega_s$ is detached from the wall, $y_0 > 0$, we can anticipate that the critical thickness for synchronization will change just as the scales of turbulence do, in particular the vertical size of the structures since $\Omega_s$ will continue to exclude all the horizontal scales.  
In these tests, while the true vorticity flux at the wall becomes part of the observed region $\Omega_f$, every new value of $y_0$ is associated with a shift in the dominant balance of turbulence dissipation, production and transport.  
In addition, the scaling of turbulence shifts from viscous units for $y^+ < 100$ \citep{Lee_moser_2015} to inertia-dominated for the large scales in the outer part \citep{Adrian2007hairpin,Smits2011}.
In order to determine the impact of $y_0$ on synchronization, we focus on two Reynolds numbers that provide sufficiently extended log regions, $Re_{\tau}=\{590, 1000\}$.  
The lower boundary of $\Omega_s$ is gradually increased from $y_0^+ = 0$ up to $y_0^+ = 100$, which corresponds to removing part of the outer layer, $y_0=\{0.17, 0.10\}$ for $Re_{\tau}=\{590, 1000\}$ respectively.
Synchronization of a horizontal layer located at the channel center will also be briefly examined.

\begin{figure}
	\centering
    \includegraphics[width=\textwidth]{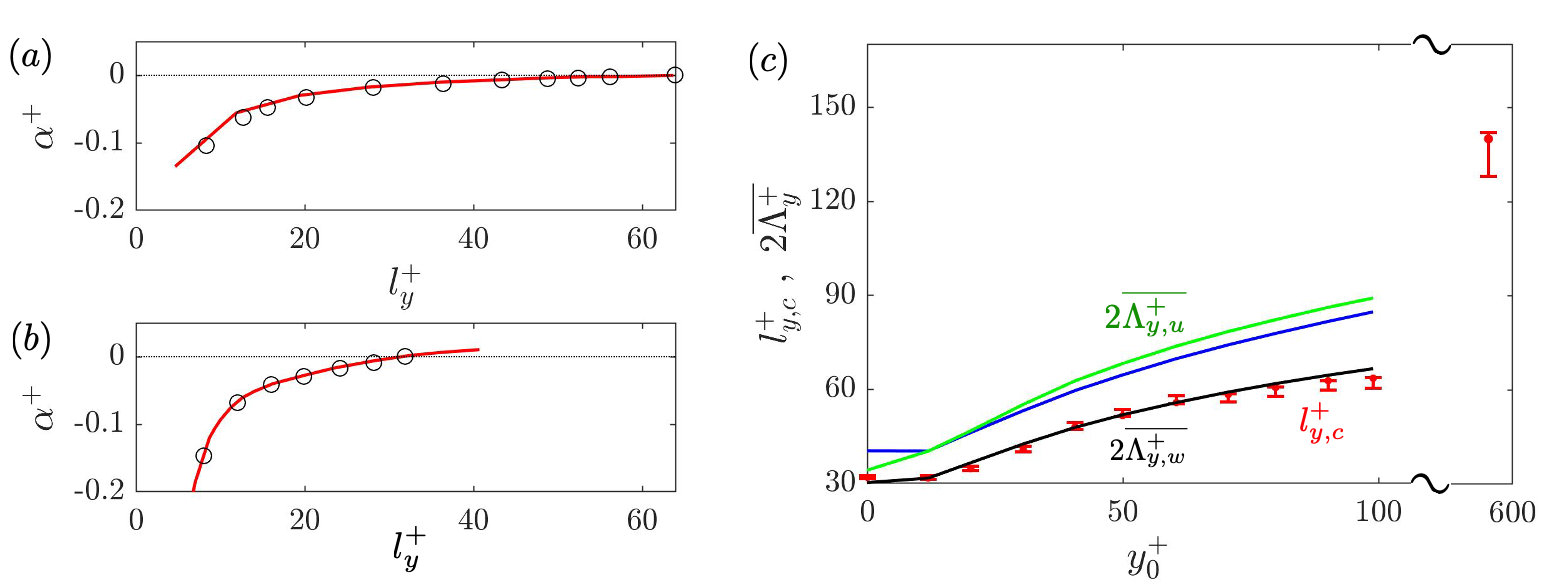}
	\caption{
		Left: Synchronization exponent with ($a$) $y_0^+ = 100$ and ($b$) $y_0^+=0$.
		Red line: $Re_{\tau}=590$; Black circles: $Re_{\tau} = 1{,}000$; horizontal dotted lines: $\alpha^+ = 0$.
		($c$) Symbols are the critical thicknesses $l_{y,c}^+$ as a function of the wall-normal distance to the cloaked layer.
		Lines are the averaged Taylor microscales based on (blue) $u$, (green) $v$, and (black) $w$ velocity components.
	}
	\label{fig:slope_log}
\end{figure}

The profile of the synchronization exponent $\alpha^+(l_y^+)$ at $y_0^+=100$ is shown in figure \ref{fig:slope_log}$a$, and is compared to the exponent when $y_0^+=0$ which is reproduced in panel $b$.
When the cloaked layer is thin, $l_y^+ \lesssim 10$, the synchronization process in the log layer is slower than within the wall layer.
As the removed region in the log layer is expanded, $\alpha^+$ monotonically increases but at a relatively weak slope.
As a result, a much thicker layer $l_{y,c}^+\approx 64$ is guaranteed to synchronize when $y_0^+ = 100$.
The synchronization exponent $\alpha^+(l_y^+)$ and the critical thickness $l_{y,c}^+$ in viscous units are unaffected by the Reynolds number (\ref{fig:slope_log}$a$,$b$), which is consistent with the inner scaling of the near-wall kinetic energy budget.
This trend suggests that synchronization of a wall-detached layer up to $y_0^+ = 100$ is still governed by similar dynamical considerations.

Starting from the equations for $\boldsymbol u_r$ and $\boldsymbol u_s$, we derive the governing equation for the squared, volume-averaged synchronization error $\mathcal E^2_{s} = \langle \| \boldsymbol u_s - \boldsymbol u_r \|^2 \rangle_{\Omega_s} \vcentcolon= \langle \| \boldsymbol e \|^2 \rangle_{\Omega_s}$, 
\begin{equation}
    \label{eq:error_energy}
    \frac{1}{2} \frac{d}{d t}\langle \| \boldsymbol e\|^2\rangle_{\Omega_s} = -\langle\boldsymbol{e}, \boldsymbol{e} \cdot \nabla \boldsymbol{u}_{r}\rangle_{\Omega_s} -\nu \langle  \| \nabla \boldsymbol e\|^2\rangle_{\Omega_s}+B,
\end{equation}
where the notation $\langle \boldsymbol a, \boldsymbol b \rangle_{\Omega_s} = \langle a_i b_i\rangle_{\Omega_s}$ is the inner product between the two vectors, volume averaged over $\Omega_s$.
The first two terms on the right-hand side of equation (\ref{eq:error_energy}) quantify the production and viscous dissipation of synchronization errors.
The last term $B$ is due to the non-zero divergence of $\boldsymbol e$ on $\partial \Omega_s$, i.e.~on the boundaries of the unobserved region $\Omega_s$.
Since $B$ only contains surface integration (see appendix \ref{sec:error_eqn} for details), it is generally much smaller than the production and dissipation terms.
Similar equations have been derived by \cite{Henshaw2003numerical} for isotropic turbulence and adopted to estimate the critical cut-off wavenumber.
Normalizing equation (\ref{eq:error_energy}) by $\mathcal E_s^2$ and assuming $\mathcal E_s = A \exp{(\alpha t)}$, the equation for exponent $\alpha$ is,

\begin{equation}
    \label{eq:error_alpha}
    \alpha = \frac{1}{\langle  \| \boldsymbol e\|^2\rangle_{\Omega_s}} \Big( -\langle\boldsymbol{e}, \boldsymbol{e} \cdot \nabla \boldsymbol{u}_{r}\rangle_{\Omega_s} - \nu \langle  \| \nabla \boldsymbol e\|^2\rangle_{\Omega_s} + B \Big).
\end{equation}

The qualitative behavior of the synchronization exponent and critical thickness in figures \ref{fig:slope_log}($a$-$b$) can be understood by aid of equation (\ref{eq:error_alpha}).
For a thin layer ($l_y \rightarrow 0$), viscous dissipation of the errors dominates the balance.
In this case, the synchronization rate is set by the action of small eddies at the Kolmogorov lengthscale. 
Since these eddies are smaller near the wall than in the outer region, synchronization occurs at a faster rate at $y^+_0 = 0$ (figure \ref{fig:slope_log}$b$) than at $y^+_0 = 100$ (figure \ref{fig:slope_log}$a$).
As a thicker layer of fluid is removed, the production rate of errors increases and eventually exceeds dissipation.
Due to a stronger mean shear in the near-wall region, the production of synchronization errors grows faster with $l_y$, which leads to a smaller critical thickness when the removed layer is closer to the wall.

The dependence of the critical thickness $l_{y,c}$ on distance from the wall is reported in figure \ref{fig:slope_log}$c$.
At each $y_0$, a bisection approach is adopted to identify the interval $[l_{y,n},l_{y,n} + \Delta y]$ (red bars in figure \ref{fig:slope_log}$c$) where the exponent $\alpha$ changes from negative to positive. 
The critical thickness (red dot) thus lies within this interval, and the value is estimated by linear interpolation using the exponents at $l_{y,n}$ and $l_{y,n} + \Delta y$.
When the lower boundary of the cloaked layer is within the range $0 \le y_0^+ \le 12$, i.e.~up to the height of peak kinetic energy production, the critical thickness for synchronization is practically unchanged. 
In the context of equation (\ref{eq:error_energy}), this behaviour is due to a comparable balance between dissipation and production of errors when the layer starts within $y_0^+ \le 12$.
However, as $y_0$ is more distant from the wall, a thicker layer of the flow can be synchronized to the reference state, or in other words the dynamics within $\Omega_s$ can be discovered from outer observations.

The trend of the critical thickness $l_{y,c}$ can also be understood with reference to the wall-normal Taylor microscales $\Lambda_{y,i}$, 
\begin{equation}
    \label{eq:taylor}
    \Lambda_{y,i} = \left(-\frac 12 \frac{d^2 \mathcal R_i}{d (\Delta y)^2} \Big|_{\Delta y=0} \right)^{-1/2} 
    \qquad \textrm{for~} i=u, v, w
\end{equation}
where $\mathcal{R}_i(\Delta y) = \langle u_i(y)u_i(y+\Delta y) \rangle_{xzt}$ is the wall-normal two-point correlation, averaged over the homogeneous horizontal directions and time. 
The Taylor microscale $\Lambda_{y,i}$ quantifies the wall-normal size of flow structures, and has recently been related to the domain of sensitivity of a velocity observations \citep{Wang2021}.
Therefore, we can use the Taylor microscales at $y_0$ and $y_0 + l_y$, 
\begin{equation}
    \label{eq:taylormean}
    2 \overline{\Lambda_{y,i}} = \Lambda_{y,i}(y_0) + \Lambda_{y,i}(y_0+l_y),
\end{equation}
as an estimate the thickness of cloaked layer that is entirely within the domain of sensitivity of the outer observations.  
The agreement between $2\overline{\Lambda_{y,w}^+}$ and $l_{y,c}^+$ in figure \ref{fig:slope_log}$c$ suggests that the critical thickness is indeed proportional to the domain of sensitivity of the available turbulence data. 
A physical interpretation of the dependence on the Taylor microscale is provided, with reference to its definition in isotropic turbulence,
\begin{equation}
    \label{eq:iso}
    \Lambda = \sqrt{15 \frac{\nu}{\mathcal D}}~u^{\prime}_{rms},
\end{equation}
where $\mathcal D$ is the dissipation rate of turbulent kinetic energy.
The square-root term in (\ref{eq:iso}) is proportional to the Kolmogorov timescale $\tau_{\eta} = \sqrt{\nu/\mathcal D}$.  Therefore, physically, the Taylor microscale is a measure of the distance swept by Kolmogorov eddies during their lifetime, while advected by the root-mean-squared velocity fluctuations. 
As long as the flow data down to the Kolmogorov eddies can be swept from $\partial \Omega_s$ throughout the cloaked region, prior to their dissipation, synchronization is guaranteed.
In turbulent channel flow, the lifetime of Kolmogorov eddies and the swept height increase with distance from the wall and, as a result, a thicker wall-normal layer can be decoded from outer observations.  

The previous arguments remain applicable in the bulk region, which is evidenced by additional synchronization tests for a cloaked horizontal layer located at the channel center, $\Omega_s = \{\boldsymbol x \in [0,L_x] \times [1 - l_y/2,1+l_y/2] \times [0,L_z]\}$.
The identified critical thickness is $l_{y,c}^+ \approx 140$, for $Re_{\tau}=590$, which is comparable to the twice Taylor microscale $2(\Lambda_{y,u},\Lambda_{y,v},\Lambda_{y,w}) \approx (139,163,110)$ at the boundary $y = 1 - l_{y,c}/2$.
It is noteworthy that $l_{y,c}^+ \approx 140$ is on the same order of magnitude as the previously reported criterion for synchronization in homogeneous isotropic turbulence, $k_c^+ \eta^+ = 0.2$ or equivalently $\Delta_c^+ = 20 \eta^+ \approx 97$ at $y = 1 - l_{y,c}/2$ \citep{Eyink2013}. Despite this similarity, it is important to recall that the interpretation of the two criteria are fundamentally different. 
The results for isotropic turbulence are in terms of a critical wavelength in Fourier space.
Specifically, the small-scale turbulence below $\Delta_c^+$ can be accurately reconstructed by enforcing all the larger scales in Fourier space.
In contrast, in our configuration, the cloaked layer removes all the streamwise and spanwise scales, in addition to the thickness $l_{y,c}^+ \approx 140$ in the wall-normal direction;  The present results then demonstrate that all the missing scales can be synchronized from the outer observations.

\begin{figure}
	\centering
	\includegraphics[width=\textwidth]{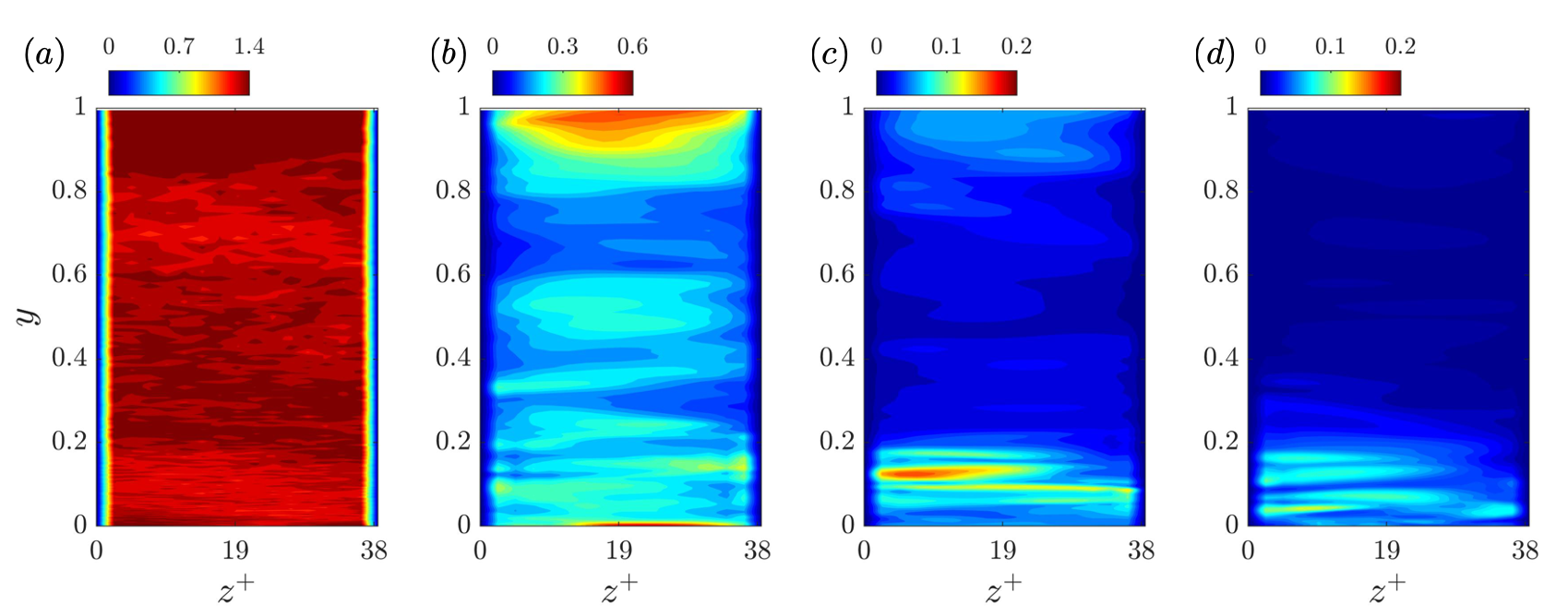}
	\caption{
		Synchronization of a vertical flow-parallel layer at $Re={590}$, when the layer width is $l_z^+ = 38$.  
		The contours show the streamwise-averaged synchronization error, normalized by the local true root-mean-squared fluctuations $\mathcal {E}_{x}(u) / u^{\prime}_{rms}$. 
		($a$-$d$) $t^+ = \{0, 64, 191, 318 \}$.
	}
	\label{fig:spanwise_layer}
\end{figure}

\subsection{Synchronization of a vertical layer}
\label{sec:vertical}
The discussion thus far has only addressed synchronization in horizontal layers, and it is expected that the orientation of the cloaked region impacts the synchronization process.
Specifically, when the layer is normal to the wall and spans the height of the channel, a distinction must be made between flow-parallel and cross-flow regions because of the effect of mean-flow advection in the latter configuration. In addition, the correlation lengthscales of the turbulent structures are different in both directions.  What remains common, however, is that the cloaked regions in both scenarios include the near-wall and outer dynamics and scales of turbulence.  In this section, we will examine synchronization in vertical layers that are oriented along either direction, at Reynolds number $Re_{\tau} = 590$.


We first consider a cloaked vertical slab that spans the height of the channel, is parallel to the flow direction and has spanwise width $l_z$, such that $\Omega_s = \{\boldsymbol x \in [0,L_x] \times [0,L_y] \times [z_0,z_0 + l_z] \}$.
The temporal evolution of streamwise-averaged synchronization error is shown in figure \ref{fig:spanwise_layer}, when $l_z^+ = 38$. 
Although the initial error is uniformly $\sqrt{2}$ times the local rms fluctuations at different wall-normal locations (panel $a$), the error decreases more rapidly in the bulk than in the near-wall region (panel $b$-$d$). 
The dominance of near-wall synchronization error in panel $d$ persists until synchronization is achieved.
The inhomogeneous evolution of errors shown in figure \ref{fig:spanwise_layer} can be explained with reference to the significant variation in the Taylor microscales in the wall-normal direction.
The spanwise Taylor microscale ranges from $2 (\Lambda_{z,u}^+, \Lambda_{z,v}^+, \Lambda_{z,w}^+) \approx (39, 22, 47)$ at the location of peak turbulence kinetic energy production, $y^+ = 12$, to $2 (\Lambda_{z,u}^+, \Lambda_{z,v}^+, \Lambda_{z,w}^+) \approx (132,114,155)$ at the channel center.
As discussed at the end of \S\ref{sec:log_layer}, the smaller Taylor microscale near the wall is indicative of a shorter swept distance during the short lifetimes of the local Kolmogorov eddies.
As a result, it is more difficult to synchronize the near-wall flow from the boundary observations.
As such, the critical width for which synchronization is guaranteed is expected to be dictated by the near-wall dynamics. 
This view is reinforced by the identified critical width $l_{z,c}^+ \approx 45$, which is the order of the Taylor microscales in the near-wall region.
The criterion $l_{z,c}^+ \lesssim 45$ ensures that the velocity observations on the spanwise boundaries can influence the entire layer during the lifetimes of the shortest surviving Kolmogorov eddies.

\begin{figure}
	\centering
	\includegraphics[width=\textwidth]{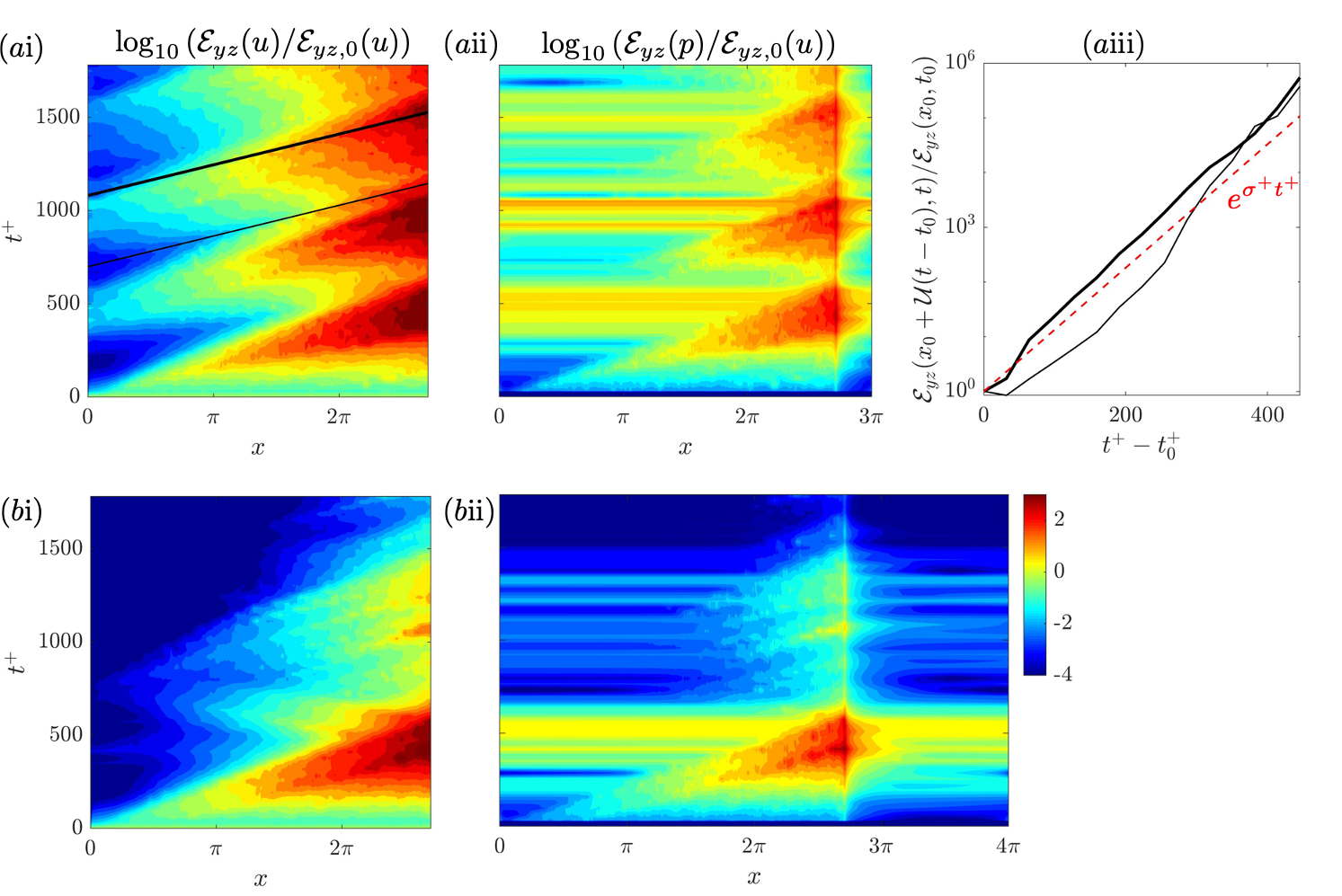}
	\caption{
		Synchronization of a vertical, crossflow layer at $Re={590}$.  
		The streamwise extent of the layer is $l_x^+ = 2.7\pi$, and two channel lengths are considered: ($a$) $L_x = 3\pi$; ($b$) $L_x = 4\pi$.  
		($a$i, $b$i) Space-time evolution of the synchronization errors in the streamwise velocity, averaged in $(y, z)$ and normalized by the initial value, $\log_{10}\left(\mathcal{E}_{yz}(u) / \mathcal{E}_{yz,0}(u)\right)$. 
		($a$ii, $b$ii) Space-time evolution of error in pressure $\log_{10}\left(\mathcal{E}_{yz}(p) / \mathcal{E}_{yz,0}(u)\right)$.
		($a$iii) Time dependence of (black) the errors along the lines $x = x_0 + \mathcal{U} (t - t_0)$ and (red dashed) Lyapunov amplification.
	}
	\label{fig:streamwise_layer_xt}
\end{figure}


Synchronization in a vertical, cross-flow layer, $\Omega_s = \{\boldsymbol x \in [x_0,x_0+l_x] \times [0,L_y] \times [0,L_z] \}$, is fundamentally different from the previous configurations due to the effect of mean advection.
Intuition suggests that it is sufficient to prescribe a single cross-flow plane of observations, akin to a simulations of boundary-layer flows, and the downstream evolution from perfect inflow data should reproduce the trajectories of the reference simulation to within machine precision. If the observations are from a physical system or an independent computational algorithm, the trajectories from the synchronization simulation are anticipated to diverge downstream at an exponential Lyapunov rate due to the chaotic nature of turbulence. 
While this intuition is helpful, it must be refined for the present channel-flow configuration because the system is closed, unlike parabolic boundary-layer flows.  


We performed a number of synchronization simulations with different cloaked streamwise extents, $l_x$.
Below a critical $l_{x,c}$ the flow synchronizes, and above it the flow trajectories diverge from the reference simulation.  The most instructive case is reported in figure \ref{fig:streamwise_layer_xt}$a$(i-iii), with $l_x = 2.7\pi$ and the domain length is $L_x=3\pi$.  The errors within the cloaked region undergo cycles of amplification and decay, and the synchronization exponent undulates around zero.  The first panel \ref{fig:streamwise_layer_xt}$a$(i) shows contours of the cross-flow-averaged synchronization error $\mathcal{E}_{yz}(u)$ as a function of streamwise distance and time.  The pattern of the contours is due to two effects:  Firstly, the initial errors which are uniform within the cloaked region amplify exponentially as the flow advects downstream (slanted lines in the $(x,t)$ plane, emanating at $t=0$). Secondly, an influx of accurate velocity data (blue region near $x=0$) reduces the magnitude of the errors within the volume. The outcome is an alternation of high- and then low-magnitude errors near $x=l_x$, which result in an alternating pattern of high and low inflow errors at $x=0$, and the process repeats. The exponential amplification of errors in shown in panel $a$(iii), along lines of constant speed $\mathcal{U}=0.61$.  The impact of the errors at $x=l_x$ on $x=0$ is despite the direct substitution $\boldsymbol{u}_s=\boldsymbol{u}_r$ within $\Omega_f$, because the pressure is not observed.  The pressure errors are reported in panel $a$(ii) versus $x\in[0,L_x]$, and indeed its elliptic nature is evident communicating the errors across the entire length of the channel. These results suggest that a larger simulation domain, specifically $\Omega_f$, could promote synchronization for this value of $l_x=2.7\pi$.  Panels $b$(i-ii) report on extending the channel from $L_x=3\pi$ to $L_x=3\pi$, and repeating the synchronization experiment.  Indeed the cloaked layer now becomes stable and synchronizes to the reference system, both in terms of the velocity (panel $b$i) and pressure (panel $b$ii).

For a closed system, when the domain size is fixed, for example circular Couette flow, increasing $l_{x}$ simultaneous reduces the forcing region $\Omega_f$ and hence the critical $l_{x,c}$ is unambiguous.  For the channel configuration, the domain length can impact the critical length for synchronization $l_{x,c}$. Nonetheless, the results showed that $l_{x,c}$ is much larger than the Taylor microscale, and instead scales with the distance travelled by the inflow during Lyapunov timescale, $l_{x,c} \sim O(\mathcal{U} \tau_{\sigma})$ where $\mathcal{U}$ is the advection velocity and $\tau_{\sigma}$ is the Lyapunov timescale.  This criterion can also be interpreted in terms of the inverse cascade of errors \citep{Leith1972predictability,Boffetta2001cascade}:
Within $\Omega_s$, perturbations at the Kolmogorov scale $\eta$ will contaminate the dynamics at the larger scale $\ell \sim 2\eta$, and the expected time of this process is proportional to the Kolmogorov timescale $\tau_{\eta}$;
Likewise, errors at $\ell \sim 2\eta$ will be transported towards the the next larger scale $\ell \sim 4\eta$, and so forth. 
The summation of eddy turn-over times at all scales is finite, and approximately equal to the Lyapunov timescale $\tau_{\sigma}$ \citep{Berera2018cascade}.
In other words, the criterion $l_x \lesssim O(\mathcal{U} \tau_{\sigma})$ prevents the small-scale errors from contaminating the large scales before the flow reaches the downstream boundary of $\Omega_s$. 
The impact of the periodic boundary condition can be lessened by extending the forcing region $\Omega_f$, and in the limit of very long $L_x$ the errors at $x=l_x$ inappreciably influence the inflow at $x=0$.  The setup then resembles predicting downstream evolution from accurate upstream data, and any infinitesimal errors due to difference in numerical setup still amplify due to chaos, and hence trajectories will diverge relative to the true system at the Luapunov rate.

\begin{figure}
	\centering
	\includegraphics[width=\textwidth]{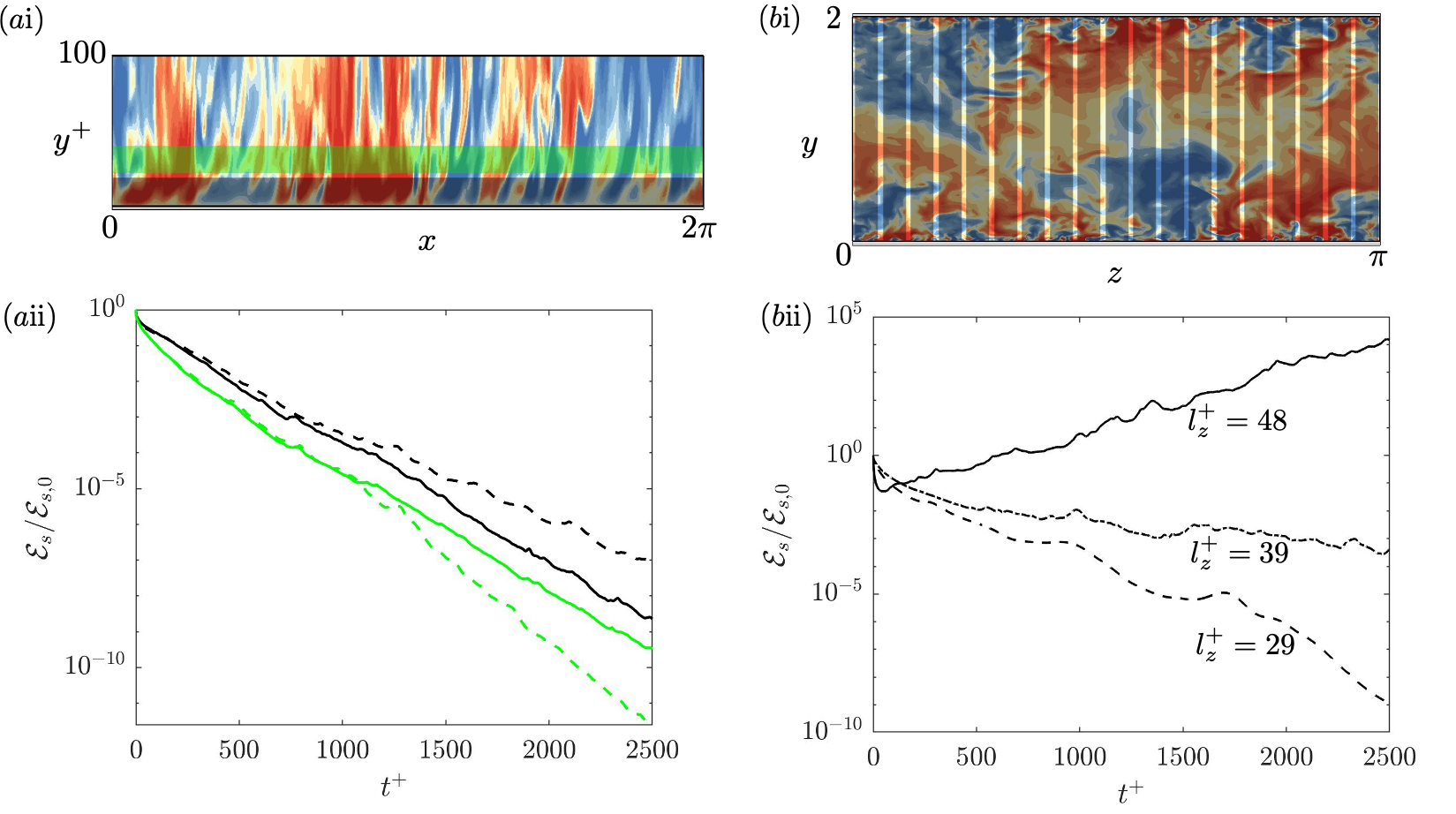}
	\caption{
		Synchronization of ($a$) two simultaneously cloaked horizontal layers and ($b$) multiple vertical flow-parallel layers, at $Re_{\tau}=590$. 
		($a$i) Schematic of the two unobserved horizontal layers. The lower layer is attached at the wall, and separated from the upper layer by one observation plane.
		($a$ii) Temporal evolution of the volume-averaged synchronization errors. 
		(Dashed lines) Only one horizontal layer is cloaked: (black) $\Omega_s  = \Omega_{s,1}$; (green) $\Omega_s = \Omega_{s,2}$. 
		(Solid lines) Both layers are simultaneously cloaked, and errors are averaged within (black) $\Omega_{s,1}$ and (green) $\Omega_{s,2}$. 
		($b$i) Schematic of the cloaked, vertical, flow-parallel layers separated by one observation plane.
		($b$ii) Temporal evolution of the volume-averaged synchronization errors for (Dashed line) $l_{z}^+ = 29$, (dashed dot line) $l_{z}^+ = 39$, and (solid line) $l_{z}^+ = 48$.
	}
	\label{fig:multiple_layer}
\end{figure}

\subsection{Synchronization of multiple layers}
\label{sec:multiple}

In the previous sections, only one layer was cloaked and the rest of the domain was observed in order to achieve synchronization. 
An intuitive inquiry is the feasibility of synchronization in two or multiple layers that are individually below their respective critical thicknesses, and that are separated from each other by some observations.
In this section, we will examine synchronization in two horizontal layers, and then proceed to explore synchronization in multiple vertical, flow-parallel layers spaced along the spanwise directions.

A schematic for simultaneously removing two horizontal layers $\Omega_{s,1} = \{\boldsymbol {x}^+ \in [0,L_x^+] \times [0,28] \times [0,L_z^+]\}$ and $\Omega_{s,2} = \{\boldsymbol {x}^+ \in [0,L_x^+] \times [29,57] \times [0,L_z^+]\}$ is shown in figure \ref{fig:multiple_layer}$a$i.  We recall that these two layers, if removed independently, have different synchronization exponents that we can denote $\alpha_1$ and $\alpha_2$.
The synchronization errors within $\Omega_{s,1}$ and $\Omega_{s,2}$ are plotted in panel \ref{fig:multiple_layer}$a$ii when both layers are simultaneously cloaked (solid lines) and are compared to the previous experiments where only one of the layers was removed (dashed lines).  The first point to note is that synchronization is successful, even though only one plane of velocity data is prescribed between the two regions.  
Secondly, beyond an initial transient, the error-decay rates in both layers are similar, and lie between $\alpha_1$ and $\alpha_2$. This behaviour of the errors demonstrates the co-dependence of the two layers: they do not synchronize independently of one another due in part to incompressibility.  
Since the two layers share the same decay rate, we can apply equation (\ref{eq:error_energy}) to the synchronization error averaged over two layers,
\begin{equation}
    \label{eq:error_energy_2layer}
    \frac{1}{2} \frac{d}{d t}\langle \| \boldsymbol e\|^2\rangle_{\Omega_s}  = \sum_{i=1,2} -\frac{\Omega_{s,i}}{\Omega_s} \left( \langle\boldsymbol{e}, \boldsymbol{e} \cdot \nabla \boldsymbol{u}_{r}\rangle_{\Omega_{s,i}}
    -\nu \langle  \| \nabla \boldsymbol e\|^2\rangle_{\Omega_{s,i}} + B_i \right).
\end{equation}
Since the production and dissipation terms include the contribution from both layers, the asymptotic decay rate is bounded by $\alpha_1$ and $\alpha_2$.

For synchronization in multiple layers, we examine cloaking of vertical slabs that span the height of the channel and are parallel to the flow direction, separated by observation data in the span (figure \ref{fig:multiple_layer}$b$i).
The thickness of each unknown layer is $l_z$, and only one plane of observation data $\boldsymbol{u}_r$ is enforced between adjacent layers. Although we set the starting location of the first layer at $z=0$, this parameter does not affect synchronization due to the statistical homogeneity of turbulence in the span.
The computational setup is slightly different from previous sections:
For both the reference and synchronization simulations, the spanwise resolution is refined to $\Delta z^+ = 2.4$ to ensure that each layer is well resolved.
In addition, the direct substitution step $\boldsymbol{u}_s(t+\Delta t)=\boldsymbol{u}_r(t+\Delta t)$ in algorithm \ref{alg:ds} is replaced by a relaxation equation $\boldsymbol u_s(t+\Delta t) \leftarrow (1 - \omega)\boldsymbol u_r(t+\Delta t) + \omega \boldsymbol u_s(t+\Delta t)$ with $\omega = 0.5$, in order to ensure stability of the synchronization simulation. 
Additional numerical experiments using different relaxation factors, down to $\omega = 0.2$, were performed and the synchronization exponents remain essentially unchanged.

The evolution of synchronization errors is reported in figure \ref{fig:multiple_layer}$b$ii for $l_z^+ = \{29, 39, 48\}$.
Similar to synchronization in horizontal layers, the errors in figure \ref{fig:multiple_layer}$b$ii decay more slowly as the removed layers become thicker. 
The critical width below which synchronization in all the layers is guaranteed is $l_{z,c}^+ \approx 42$, which is slightly smaller than the criterion for removing a single layer, $l_{z,c}^+ \approx 45$ (c.f. \S\ref{sec:vertical}).  Since the balance of production and dissipation of errors per unit width is statistically stationary in the span, the difference in the critical width is due to the increase in the number of interfaces.  Specifically, the cumulative contribution from the boundary terms in equation (\ref{eq:error_energy_2layer}) increases as the number of interfaces between $\Omega_s$ and $\Omega_f$ increases.

This configuration can also be discussed in relation to earlier studies of synchronization in spectral space.  
The availability of data at a constant spanwise intervals is comparable to observing spanwise wavenumbers $k_z^+ \le 2 \pi / 2 l_z^+$. 
Successful synchronization below $l_{z,c}^+=42$, or equivalently above $k_{z,c}^+=0.0748$,  is not straightforward to expressed in terms of the Kolmogorov lengthscale which depends on wall-normal distance. 
For the benefit of this discussion, we remark the Kolmogorov scale is approximately $\eta^+=1.66$ at the location of peak turbulence-kinetic-energy production, where the Taylor microscale was commensurate with $l_{z,c}^+$.  Therefore $l_{z,c}^+\approx 25 \eta^+$, or $k_{z,c} \eta \approx 0.12 $, similar to the criterion for synchronization in spectral space \citep{Yoshida2005,Eyink2013}.  Despite this favorable comparison, we note that the present configuration is different, not least because all the streamwie and wall-normal wavenumbers are unknown within the cloaked region.  
More importantly, within the cloaked region, the present configuration features dynamics that are unique to wall-bounded turbulence that must be synchronized within the cloaked region (e.g.\,wall-vorticity flux, wall-normal separation of dissipation and production, wall-normal dependence of ejections and sweeps,...etc).  

The success of synchronization of multiple layers, with merely planar observations in between layers, raises an interesting question: Given planar observations, is it possible to accurately reconstruct the velocity field in a sub-domain bounded by that surface without simulating the entire system?
This question is addressed in the next section.


\subsection{Synchronization in sub-domain simulations}
\label{sec:subdomain}

\begin{figure}
	\centering
	\includegraphics[width=\textwidth]{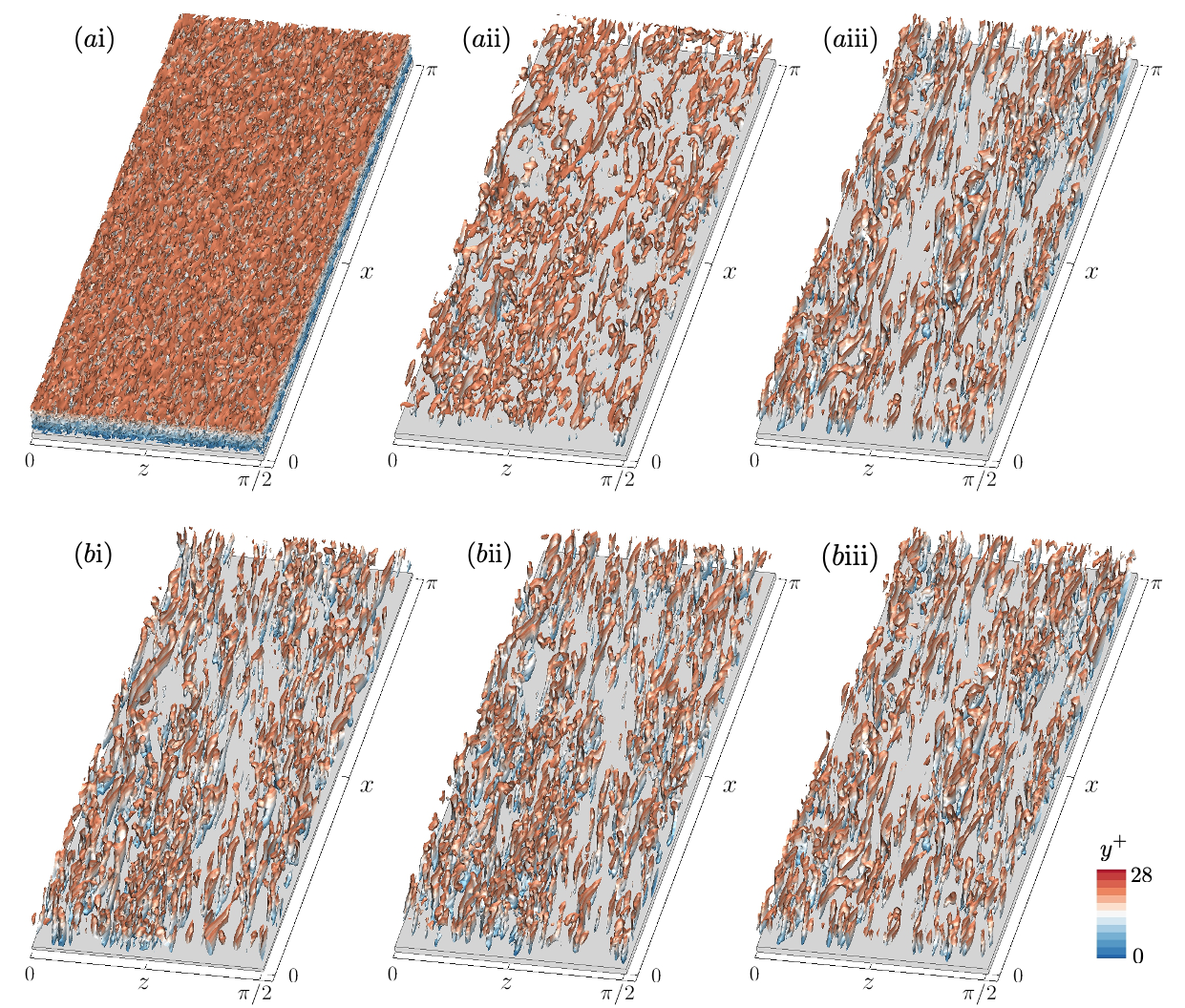}
	\caption{
	    Instantaneous vortical structures visualized using the $\lambda_2$ vortex identification criterion with threshold $\lambda_2 = -4$. 
	    ($a$) Synchronization simulation in a wall layer $l_y^+ = 28$;
	    ($b$) Reference simulation.  
	    (i-iii) $t^+= \{0, 13, 1000\}$.
	}
	\label{fig:subdomain_vort}
\end{figure}

A large volume of velocity observations could be difficult to acquire and, in practice, is unlikely to span the entire system beyond $\Omega_s$.  Even if observations are available for the complete state outside $\Omega_s$, the computational cost of synchronization simulations in the full system can be prohibitive at high Reynolds numbers. It is therefore desirable to attempt synchronization of the cloaked region by simulating a sub-domain of the full system.  
An example using Kolmogorov flow is provided in Appendix \ref{sec:Kflow}, where we perform synchronization simulations in a sub-volume using velocity observations on the bounding planes only.  Here we retain our focus on channel flow, and return to the case of synchronization of a wall-attached layer but with an observer system that is a sub-volume of the reference configuration and more limited observations.

The reference system $\boldsymbol{u}_r$ is still turbulent flow in the entire channel, and the velocity field is observed at one horizontal plane only, $\Omega_f = \{ y = \tilde l_y \}$.
The synchronization simulation $\tilde {\boldsymbol u}_s$ is performed in a wall-attached layer $\Omega = \{y \in [0,\tilde l_y] \}$, where the observed velocity is enforced at $y=\tilde{l}_y$ along with a homogeneous Neumann condition on the pressure because $p$ is not observed. 
Together with periodic boundary conditions in the horizontal directions and no-slip at the wall, these six faces bound $\Omega_s$, without any additional data.
The initial estimate of $\boldsymbol{u}_s$ is either the mean flow superposed with white noise or the slightly disturbed true state, as explained in \S\ref{sec:methods}.
Our objective is to synchronize the flow in $\Omega_s = \{\boldsymbol x \in [0,L_x] \times [0,\tilde l_y) \times [0,L_z]\}$ to the reference state $\boldsymbol u_r$.

The subdomain synchronization process at $Re_{\tau}=590$ with $\tilde l_y^+ = 28$ is visualized in figure \ref{fig:subdomain_vort}$a$, and is compared with the true state in panels $b$.
The white noise at the initial time (panel $a$i) decays quickly and the vortical structures near the top boundary (panel $a$ii) become similar to the true field, although the wall-attached vorticies are absent.
After a sufficiently long time, all of $\Omega_s$ is affected by the top boundary condition and the estimated state within the sub-domain simulation synchronizes to the reference state:  all the scales and structures in panel $a$iii identically match those in $b$iii, to within machine precision.
These qualitative properties are similar to the full-domain synchronization results in figure \ref{fig:quality}, although here the observer system is only a sub-domain simulation with height $l_y^+=28$.  

\begin{figure}
	\centering
	\includegraphics[width=\textwidth]{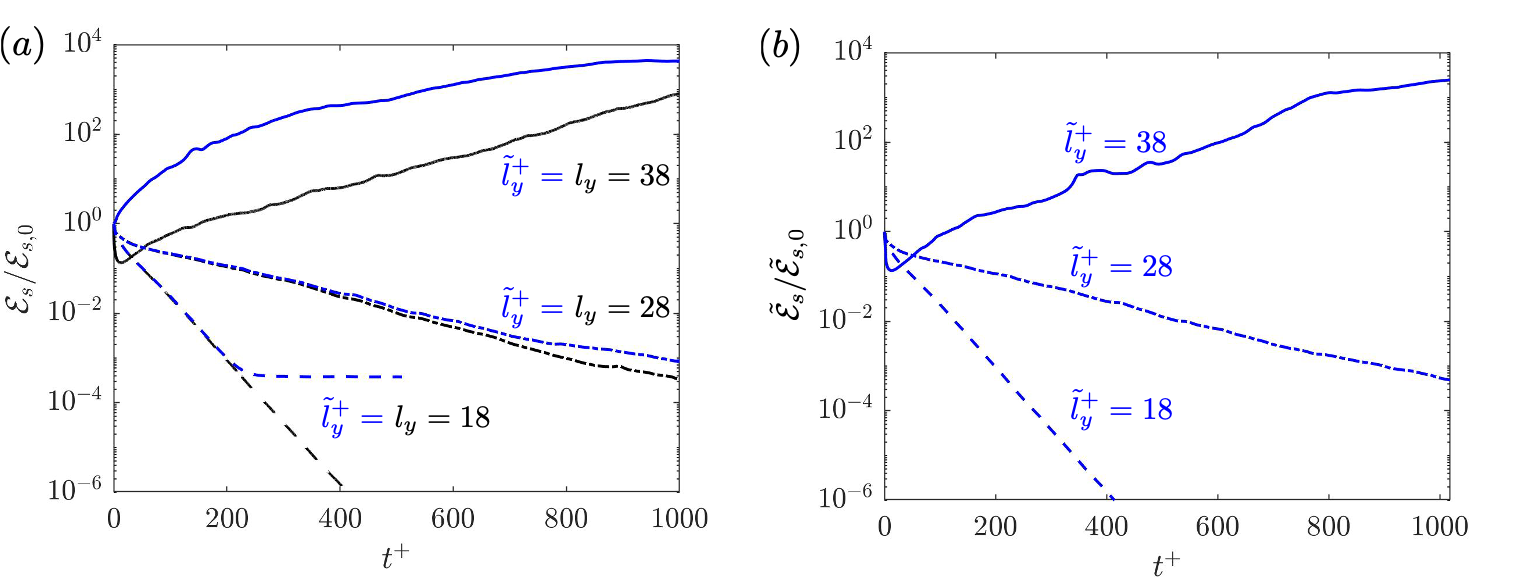}
	\caption{
		Synchronization in a sub-domain of the channel, at $Re_{\tau}=590$. 
		($a$) Volume-averaged errors $\mathcal E_{s}$ of the instantaneous velocity, when synchronization simulations are performed in (blue) a sub-domain, compared to (black) the full domain. 
		The reference velocity $\boldsymbol u_r$ in the definition of the errors was obtained from a full-domain simulation.
		($b$) The errors of the sub-domain synchronization experiments are re-evaluated using a reference velocity from a simulation in the same subdomain, $\tilde {\mathcal E}_{s} = \langle \|\tilde{\boldsymbol u}_s - \tilde{\boldsymbol u}_r \|^2\rangle_{\Omega_s}^{1/2}$.
		(Dashed lines) $  \tilde l_y^+ =l_y^+ = 18$; (dashed dot lines) $ \tilde l_y^+=  l_y^+=28$; (solid lines) $ \tilde  l_y^+= l_y^+=38$.
	}
	\label{fig:subdomain_error}
\end{figure}

The errors from the sub-domain synchronization are reported in figure \ref{fig:subdomain_error}$a$ (blue lines), where they are compared to results using the approach from \S\ref{sec:wall_layer} for full-domain synchronization (black lines).
We verified that the critical layer thickness remains unchanged, $\tilde{l}^+_{y,c} = l^+_{y,c} \approx 32$. 
When the sub-domain height is larger than this value, $\tilde{l}_y > l_{y,c}$, the errors diverge similar to the previously report behaviour in \S\ref{sec:wall_layer}.  
For smaller layer heights, below the critical value, the errors decay exponentially which indicates convergence towards the true flow trajectories, initially at the same rate as the full-system configuration.  
However, a significant difference can be observed:
The errors in the present sub-domain approach do not decay to machine precision, but rather saturate at approximately $10^{-3}$ of the initial value, or equivalently three orders of magnitude lower than the reference r.m.s. fluctuations. 
These persistent long-time errors are not due to failure of synchronization of the dynamics as described by the Navier-Stokes equations, but are rather due to the differences in the numerical solutions of the governing equations in a sub-domain versus the full channel, specifically the fractional step method.  In support of this argument, we performed a new reference simulation within the wall-attached layer starting from the true initial condition, which we term $\tilde{\boldsymbol u}_r$.  This computation does not match the reference solution of the full system $\boldsymbol{u}_r$, and the difference is on the order of $10^{-3}$\textemdash a matter that we address below. More importantly, the synchronization solution $\tilde{\boldsymbol u}_s$ from an arbitrary initial condition converges to $\tilde{\boldsymbol u}_r$ to within machine precision (figure \ref{fig:subdomain_error}$b$); the sub-system is therefore Lyapunov stable and can accurately synchronize when $\tilde{l}_{y} < \tilde{l}_{y,c} = l_{y,c}$.  

The reason that the reference simulations in the sub- and full domains, $\tilde{\boldsymbol u}_r$ and $\boldsymbol u_r$, differ can be explained using a similar analysis as \cite{Wu2020resimulation}.
Briefly, in addition to the same initial condition, the intermediate velocity of the fractional-step algorithm for solving the Navier-Stokes equations must be enforced at $y = \tilde l_y$ to eliminate the mismatch.  This velocity is not physical, and is only introduced for the purpose of numerically solving the equations. We have demonstrated synchronization to $\tilde{\boldsymbol u}_r$ in the sub-system, and the deviation from the full system is a matter of a difference in the numerical model.  
Notwithstanding this technical detail, the results presented here and in Appendix \ref{sec:Kflow} demonstrate that with only planar observations, synchronization in a sub-domain that satisfies the critical size criteria can accurately converge onto the true trajectories of the full system, to within the errors of the numerical model (figures \ref{fig:subdomain_error}$a$-$b$), and hence is sufficiently accurate to evaluate any flow quantity of interest including velocity gradients or vortical structures (c.f. figure \ref{fig:subdomain_vort}).  

\section{Conclusions}
\label{sec:conclusion}
Synchronization of chaos in turbulent channel flow was attempted using continuous data assimilation techniques.
Fully-resolved observations of the turbulence are available outside a cloaked, or unobserved, region of the flow.  These observations are directly substituted in the synchronization simulation at every time step in order to drive the missing flow towards the reference state.
The temporal evolution of the estimation error was adopted to evaluate a synchronization exponent, which determines the success and rate of synchronization.
When synchronization is successful, all the turbulence scales in the cloaked region are accurately reestablished, and the synchronization exponent is independent of the initial condition. 

Synchronization was examined in detail for a cloaked, wall-attached horizontal layer.
After the initial transient, a successful synchronization features monotonic decay of the errors within the layer.  The synchronization rate, when scaled in wall units, is independent of the instantaneous reference state and the Reynolds number up to $Re_{\tau} = 1000$. The critical layer thickness is $l_{y,c}^+ \approx 32$, which indicates that the near-wall turbulence, including the true vorticity source, are interpretable from outer observations. 

When the unknown horizontal layer is detached from the wall, its synchronization rate and critical thickness depend on its wall-normal distance, $y_0$.
A thin layer has a relatively slow synchronization rate compared to a wall-attached counterpart, but the critical thickness increases monotonically with $y_0$.
Up to $y_0^+ = 100$, the synchronization rate and critical thickness expressed in wall units are independent of the Reynolds number.  
These trends are the outcome of the balance in the equation governing the synchronization errors, which features a source term at the boundary of the cloaked region, production of errors against the mean shear and viscous dissipation of errors. Relative to the scales of turbulence, the critical thickness of a cloaked layer is comparable to twice the Taylor microscale which quantifies the distance swept by Kolmogorov eddies within their lifetimes as they are transported by the root-mean-square velocity.

Simultaneous synchronization in multiple volumes was investigated by considering two horizontal layers separated by a single plane of observations.  The synchronization rate was bounded by the exponents that correspond to cloaking each layer alone.
We also examined synchronization of vertical layers that are parallel to the flow, and separated by a single observation planes.  The critical layer width that ensured synchronization was identified, $l_{z,c}^+ \approx 42$, and is on the order of the near-wall Taylor microscale of the turbulence.
The criterion for synchronization of a vertical, cross-flow volume is set by the mean advection and the Lyapunov timescale, and is therefore less restrictive than the criteria in the other directions for the range of Reynolds numbers examined herein.

Our final experiments, which are of practical interest, attempted reconstruction of the cloaked region using simulations of a sub-domain of the reference system, and with limited observations.   
We revisited synchronization of a wall layer, when the available data are one plane of instantaneous velocity observed from the reference channel-flow system. The observer system is then the cloaked near-wall layer and that plane of data only.  The reconstructed flow fields perfectly matched re-simulations of the sub-domain using the true initial conditions, and both exhibited some deviation from the true, full channel-flow system due to the discretization scheme of the Navier-Stokes equations.  This mismatch can be reduced by increasing the number of observed velocity planes. In general, synchronization in a sub-domain using boundary data on all the faces is feasible, as long as the herein reported critical dimensions are observed.  All together, the present results demonstrate the potential of augmenting experimental observations using continuous data assimilation techniques that are efficient, easy to implement, and can provide non-intrusive access to true flow trajectories.

\par\bigskip
\noindent
\textbf{Funding.} This work was supported in part by the Office of Naval Research (grant N00014-20-1-2715). 
Computational resources were provided by the Maryland Advanced Research Computing Center (MARCC).

\par\bigskip
\noindent
\textbf{Declaration of interests.} 
The authors report no conflict of interest.
 
\par\bigskip
\noindent
\textbf{Author ORCIDs.} \\
Tamer A. Zaki, \url{https://orcid.org/0000-0002-1979-7748}

\appendix

\section{Equation for volume-averaged synchronization error}
\label{sec:error_eqn}

In this appendix we outline the procedures to derive the equation for the volume-averaged synchronization error, $\mathcal E_{s} = \langle \|\boldsymbol e \|^2 \rangle_{\Omega_s}^{1/2}$.
By subtracting the Navier-Stokes equations for $\boldsymbol u_s$ and $\boldsymbol u_r$, we obtain the equations governing $\boldsymbol e := \boldsymbol u_s - \boldsymbol u_r$ and $\zeta := p_s - p_r$, \begin{eqnarray}
    \label{eq:error_mom}
    \frac{\partial \boldsymbol{e}}{\partial t}+\boldsymbol{e} \cdot \nabla \boldsymbol{u}_{r}+\boldsymbol{u}_{r} \cdot \nabla \boldsymbol{e}+\boldsymbol{e} \cdot \nabla \boldsymbol{e} &=& -\nabla \zeta+\nu \nabla^{2} \boldsymbol{e} + \boldsymbol f, \\
    \label{eq:error_div}
    \nabla \cdot \boldsymbol e &=& b(\boldsymbol x_b) ~\delta(\boldsymbol{x} - \boldsymbol{x}_{b}).
\end{eqnarray}
The forcing $\boldsymbol f$ represents direct substitution of the observations in $\Omega_f$, which ensures that $\boldsymbol e = \boldsymbol 0$ for all $\boldsymbol{x} \in \Omega_f$.
Note that the pressure difference $\zeta$ is non-zero through the entire channel because the reference pressure $p_r$ is not observed.
Due to discontinuity between the observed and estimated velocity fields at all $\boldsymbol x_b \in \partial\Omega_s$, the corresponding divergence of the synchronization error is not zero, which leads to the source term $b(\boldsymbol x_b)$ in (\ref{eq:error_mom}).
The specific form of the source term depends on the numerical discretization.
By calculating the dot product of (\ref{eq:error_mom}) with $\boldsymbol e$ and then averaging over $\Omega_s$, the equation for $\mathcal E_{s}^2 = \langle \|\boldsymbol e \|^2 \rangle_{\Omega_s}$ can be derived, 
\begin{equation}
    \label{eq:error_energy_2}
    \frac{1}{2} \frac{d}{d t}\langle \| \boldsymbol e\|^2\rangle_{\Omega_s} = -\langle\boldsymbol{e}, \boldsymbol{e} \cdot \nabla \boldsymbol{u}_{r}\rangle_{\Omega_s} -\nu \langle  \| \nabla \boldsymbol e\|^2\rangle_{\Omega_s}+B. 
\end{equation}
Other terms that can be written in conservative form vanish, either by virtue of $\boldsymbol e = \boldsymbol 0$ on the boundaries of $\Omega_s$ or appropriate boundary conditions, for example periodicity in the horizontal directions for a horizontal slab and no-slip conditions on the wall.
The boundary term in the kinetic-energy equation is
\begin{equation}
    \label{eq:error_B}
    B = \frac{1}{\Omega_s} \int_{\Omega_s} b(\boldsymbol x_b) \delta (\boldsymbol x - \boldsymbol x_b) \zeta(\boldsymbol x)dV = \frac{1}{\Omega_s} \int_{\partial \Omega_s} b(\boldsymbol x_b) \zeta(\boldsymbol x_b) dS. 
\end{equation}
This term only contains a surface integration over the boundary $\partial \Omega_s$ and is generally much smaller than the production and dissipation in (\ref{eq:error_energy_2}).
For the wall-attached horizontal layer $\Omega_s = \{\boldsymbol x \in [0,L_x] \times [0,l_y] \times [0,L_z]\}$ discussed in \S\ref{sec:wall_layer}, the boundary term (\ref{eq:error_B}) becomes,
\begin{equation}
    \label{eq:error_B_WallLayer}
    B = \frac{1}{\Omega_s} \int_{\Omega_s} b(x,z) \delta (y-l_y) \zeta(x,y,z)dV = \frac{1}{\Omega_s} \int_{\partial \Omega_s} b(x,z) \zeta(x,l_y,z) dxdz. 
\end{equation}

\section{Subdomain synchronization in Kolmogorov flow}
\label{sec:Kflow}

\begin{figure}
	\centering
	\includegraphics[width=1.0\textwidth]{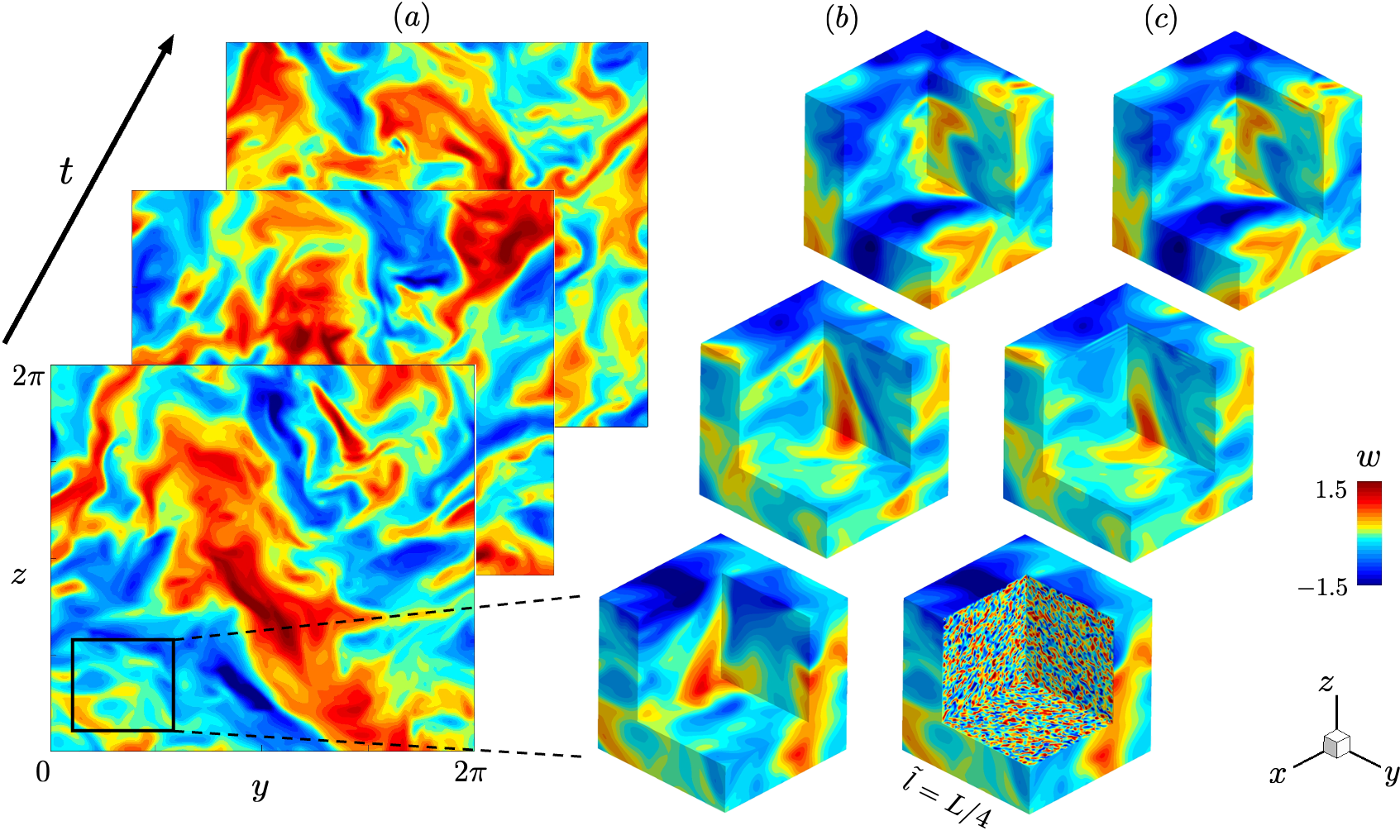}
	\caption{Instantaneous visualizations of synchronization in Kolmogorov flow.  
	($a$) Contours of the reference spanwise velocity $w_r$, plotted at $x=L/2$.
	($b$) Enlarged, three-dimensional views of the reference subdomain of interest and ($c$) of synchronization using a sub-domain simulation.
	($a$-$c$) The three snapshots from bottom to top in correspond to $t=\{0, 1, 4\}$.
	 }
	\label{fig:Kflow_visualize}
\end{figure}

\begin{figure}
	\centering
	\includegraphics[width=0.5\textwidth]{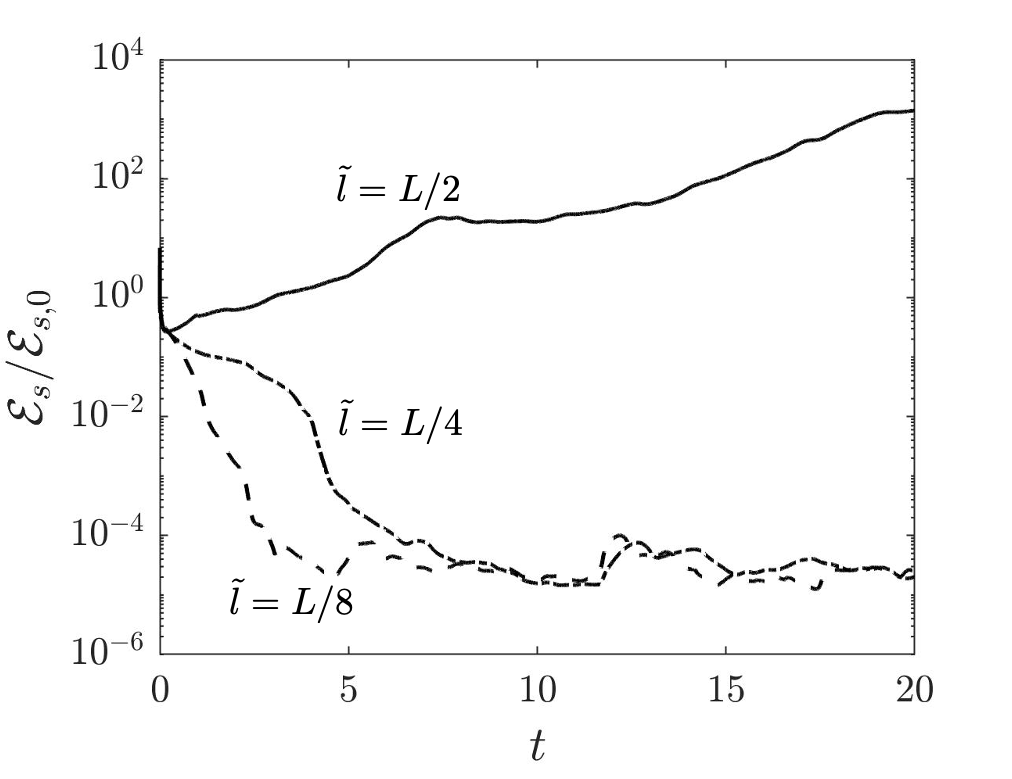}
	\caption{
    Temporal behaviour of errors in sub-domain synchronization simulations of Kolmogorov flow.  
	The volume-averaged errors $\mathcal E_{s}$ are normalized by their initial value $\mathcal E_{s,0}$.
	(Dashed) $\tilde l = L/8$; (dashed-dot) $\tilde l = L/4$; (solid) $\tilde l = L/2$.
	}
	\label{fig:Kflow_error}
\end{figure}

In \S\ref{sec:subdomain}, we demonstrated that the wall layer in turbulent channel flow can be accurately recovered by performing the synchronization simulation in a sub-domain, with the limited observations of the velocity on the top boundary only.  The synchronization sub-domain thus spanned the horizontal dimensions of the entire channel, with all the streamwise and spanwise turbulent scales being unknown.  
Motivated by practical limitations, in this appendix we consider the more specialized case where the synchronization sub-domain is a general, three-dimensional, rectangular sub-volume that is neither periodic nor attached to a boundary, and the observations are limited to the velocity data on the surrounding planes only.  The synchronization experiments are performed in Kolmogorov flow which was the subject of previous studies of synchronization, albeit solely in Fourier space \citep{Eyink2013}.

The reference simulation is performed in a tri-periodic, cubic box with sides $L = 2\pi$, and driven by a body force $\boldsymbol f_b = (0.2 \sin y, 0, 0)$.
The Reynolds number is $Re_{\Lambda} =u^{\prime}_{rms} \Lambda /\nu = 146$, where the Taylor microscale $\Lambda = \sqrt{15 \nu / \mathcal D}\, u^{\prime}_{rms}$ is evaluated using global dissipation $\mathcal D$.
The computational domain is discretized on a uniform grid with 256 points in each direction. 
The incompressible Navier-Stokes equations are solved using the same numerical algorithm as in \S\ref{sec:methods}. 
Due to periodicity, the pressure Poisson equation is solved using Fourier transform in all three directions. 
Once the flow reaches a statistically stationary state (see sample snapshots in figure \ref{fig:Kflow_visualize}$a$), instantaneous velocity data on six faces of a cubic subdomain $[l_0, l_0 + \tilde l]^3$ are extracted and stored. 

The synchronization simulations are performed in sub-domains bounded by these six faces, and the velocity observations are prescribed as Dirichlet boundary conditions.
Since pressure is not observed, Neumann boundary conditions $\partial p / \partial n = 0$ are adopted, and a Fourier-cosine transforms are performed to solve the pressure Poisson equation. 
The starting location of the sub-domain is fixed $l_0 = L/16$, and three subdomain sizes $\tilde l=\{\sfrac{1}{8}, \sfrac{1}{4}, \sfrac{1}{2}\}L$ are considered.
The initial estimate for the first two cases is white noise proportional to the root-mean-squared velocity fluctuations, and a Lyapunov-type experiment is performed for $\tilde l=L/2$.

The sub-domain synchronization simulation for $\tilde l = L/4$ is visualized in figure \ref{fig:Kflow_visualize}$c$ at $t=\{0, 1, 4\}$, and compared to the true state in panel $b$.
The velocity field on the surface of the cube is identical to the true state because the continuous data assimilation in this configuration is effectively prescribed as the time-dependent velocity boundary condition.
Deviations from the reference state occur only at interior points (see figures \ref{fig:Kflow_visualize}$b$ and $c$), and diminish as the flow evolves in time.
The volume-averaged synchronization error is plotted in figure \ref{fig:Kflow_error} (dashed-dotted line), normalized by its initial value.  The error decays to $10^{-4}$ and then saturates, similar to the behaviour discussed in \S\ref{sec:subdomain} when the reference velocity is obtained from the full-domain simulation. 
Note that the error would decay to machine precision if compared to a reference simulation in the sub-domain, starting from an accurate initial condition.

Figure \ref{fig:Kflow_error} shows the synchronization errors for all three cases, $\tilde l=\{\sfrac{1}{8}, \sfrac{1}{4}, \sfrac{1}{2}\}L$. The critical sub-domain size $\tilde{l}_c$ lies between $tilde{l}=L/4$ and $tilde{l}=L/2$.  Because the large-scale flow is inhomogeneous in $y$ direction, the local Taylor microscales are evaluated at each $y$ position from the two-point correlations.
The maximum value over $y$ and the $\{u ,v, w\}$ components are $2\Lambda_{x,u}=0.95$, $2\Lambda_{y,v}=0.82$, and $2\Lambda_{z,w}=0.75$, which are all smaller than $L/4$.
Similar to removing streamwise layers in \S\ref{sec:multiple}, the critical sub-domain size in $x$ exceeds $2\Lambda_x$ because of the mean advection effect. 
In effect, as long as the sub-domain size is below the threshold for synchronization in one direction, the flow can synchronize to the reference state by enforcing planar velocity data on the six faces.

\bibliographystyle{jfm}
\bibliography{reference}

\end{document}